\documentclass[twocolumn]{aastex631}
\usepackage{colortbl}
\shorttitle{Demographics of Star-Forming Clumps}
\shortauthors{Martin et al.}
\graphicspath{{./}{figures/}}
\begin{document}

\title{UV-Bright Star-Forming Clumps and Their Host Galaxies in UVCANDELS at 0.5 $\leq$ z $\leq$ 1}

\author[0000-0002-6632-4046]{Alec Martin}
\affiliation{Department of Physics and Astronomy, University of Missouri, Columbia, MO 65211, USA}

\author[0000-0003-2775-2002]{Yicheng Guo}
\affiliation{Department of Physics and Astronomy, University of Missouri, Columbia, MO 65211, USA}

\author[0000-0002-9373-3865]{Xin Wang}
\affiliation{School of Astronomy and Space Science, University of Chinese Academy of Sciences (UCAS), Beijing 100049, China}
\affiliation{National Astronomical Observatories, Chinese Academy of Sciences, Beijing 100101, China}
\affiliation{Institute for Frontiers in Astronomy and Astrophysics, Beijing Normal University, Beijing 102206, China}

%%%%%% Block 2 %%%%%%%%%%

\author[0000-0002-6610-2048]{Anton M. Koekemoer}
\affiliation{Space Telescope Science Institute, 3700 San Martin Dr., Baltimore, MD 21218, USA}

\author[0000-0002-9946-4731]{Marc Rafelski}
\affiliation{Space Telescope Science Institute, 3700 San Martin Dr., Baltimore, MD 21218, USA}
\affiliation{Department of Physics and Astronomy, Johns Hopkins University, Baltimore, MD 21218, USA}

\author[0000-0002-7064-5424]{Harry I. Teplitz}
\affiliation{IPAC, Mail Code 314-6, Caltech, 1200 E. California Blvd., Pasadena, CA 91125, USA}

\author[0000-0001-8156-6281]{Rogier A. Windhorst}
\affiliation{School of Earth and Space Exploration, Arizona State University, Tempe, AZ 85287-1404, USA}

%%%%%% Block 3 %%%%%%%%%%

\author[0000-0002-8630-6435]{Anahita Alavi}
\affiliation{IPAC, Mail Code 314-6, Caltech, 1200 E. California Blvd., Pasadena, CA 91125, USA}

\author[0000-0001-9440-8872]{Norman A. Grogin}
\affiliation{Space Telescope Science Institute, 3700 San Martin Dr., Baltimore, MD 21218, USA}

\author[0000-0002-0604-654X]{Laura Prichard}
\affiliation{Space Telescope Science Institute, 3700 San Martin Dr., Baltimore, MD 21218, USA}

\author[0000-0003-3759-8707]{Ben Sunnquist}
\affiliation{Space Telescope Science Institute, 3700 San Martin Dr., Baltimore, MD 21218, USA}

%%%%%% Block 4 %%%%%%%%%
\author[0000-0002-8680-248X]{Daniel Ceverino}
\affiliation{Departamento de Fisica Teorica, Modulo 8, Facultad de Ciencias, Universidad Autonoma de Madrid, 28049 Madrid, Spain}
\affiliation{CIAFF, Facultad de Ciencias, Universidad Autonoma de Madrid, 28049 Madrid, Spain}

\author[0000-0003-3691-937X]{Nima Chartab}
\affiliation{The Observatories of the Carnegie Institution for Science, 813 Santa Barbara St., Pasadena, CA 91101, USA}

\author[0000-0003-1949-7638]{Christopher J. Conselice}
\affiliation{Jodrell Bank Centre for Astrophysics, University of Manchester, Oxford Road, Manchester M13 9PL, UK}

\author[0000-0002-7928-416X]{Y. Sophia Dai}
\affiliation{Chinese Academy of Sciences South America Center for Astronomy (CASSACA), National Astronomical Observatories(NAOC),20A Datun Road, Beijing 100012, China}

\author[0000-0003-4174-0374]{Avishai Dekel}
\affiliation{Racah Institute of Physics, The Hebrew University of Jerusalem, Jerusalem 91904, Israel}

\author[0000-0003-2098-9568]{Jonathan P. Gardner}
\affiliation{Astrophysics Science Division, NASA Goddard Space Flight Center, 8800 Greenbelt Rd, Greenbelt, MD 20771, USA}

\author[0000-0003-1530-8713]{Eric Gawiser}
\affiliation{Department of Physics and Astronomy Rutgers University Piscataway, NJ 08854, USA}

\author[0000-0001-6145-5090]{Nimish P. Hathi}
\affiliation{Space Telescope Science Institute, 3700 San Martin Dr., Baltimore, MD 21218, USA}

\author[0000-0001-8587-218X]{Matthew J. Hayes}
\affiliation{Stockholm University, Department of Astronomy and Oskar Klein Centre for Cosmoparticle Physics, AlbaNova University Centre, SE-10691, Stockholm, Sweden}

\author[0000-0003-1268-5230]{Rolf A. Jansen}
\affiliation{School of Earth and Space Exploration, Arizona State University, Tempe, AZ 85287-1404, USA}

\author[0000-0001-7673-2257]{Zhiyuan Ji}
\affiliation{Steward Observatory, University of Arizona, 933 N. Cherry Avenue, Tucson, AZ 85721, USA}

\author[0000-0003-3385-6799]{David C. Koo}
\affiliation{UCO/Lick Observatory, Department of Astronomy and Astrophysics, University of California, Santa Cruz, CA, USA}
\affiliation{Department of Astronomy and Astrophysics, University of California, Santa Cruz, 1156 High Street, Santa Cruz, CA 95064, USA}

\author[0000-0003-1581-7825]{Ray A. Lucas}
\affiliation{Space Telescope Science Institute, 3700 San Martin Dr., Baltimore, MD 21218, USA}

\author[0000-0001-8057-5880]{Nir Mandelker}
\affiliation{Racah Institute of Physics, The Hebrew University of Jerusalem, Jerusalem 91904, Israel}

\author[0000-0001-7166-6035]{Vihang Mehta}
\affiliation{IPAC, Mail Code 314-6, Caltech, 1200 E. California Blvd., Pasadena, CA 91125, USA}

\author{Bahram Mobasher}
\affiliation{Department of Physics and Astronomy, University of California, Riverside, 900 University Ave, Riverside, CA 92521, USA}

\author[0000-0001-5294-8002]{Kalina V. Nedkova}
\affiliation{Department of Physics and Astronomy, Johns Hopkins University, Baltimore, MD 21218, USA}

\author{Joel Primack}
\affiliation{Department of Physics, University of California, Santa Cruz, CA, 95064, USA}
%\author[0000-0003-1581-7825]{Ray A. Lucas}
%\affiliation{Space Telescope Science Institute, 3700 San Martin Dr., Baltimore, MD 21218, USA}

\author[0000-0002-5269-6527]{Swara Ravindranath}
\affiliation{Space Telescope Science Institute, 3700 San Martin Dr., Baltimore, MD 21218, USA}

\author[0000-0002-4271-0364]{Brant E. Robertson}
\affiliation{Department of Astronomy and Astrophysics, University of California, Santa Cruz, 1156 High Street, Santa Cruz, CA 95064, USA}

\author[0000-0001-7016-5220]{Michael J. Rutkowski}
\affiliation{Department of Physics and Astronomy, Minnesota State University, Mankato, MN 56001, USA}

\author[0000-0002-0364-1159]{Zahra Sattari}
\affiliation{The Observatories of the Carnegie Institution for Science, 813 Santa Barbara St., Pasadena, CA 91101, USA}
\affiliation{Department of Physics and Astronomy, University of California, Riverside, 900 University Ave, Riverside, CA 92521, USA}

\author[0000-0002-2390-0584]{Emmaris Soto}
\affiliation{Computational Physics, Inc., Springfield, VA, USA}

\author[0000-0003-3466-035X]{{L. Y. Aaron} {Yung}}
\affiliation{Astrophysics Science Division, NASA Goddard Space Flight Center, 8800 Greenbelt Rd, Greenbelt, MD 20771, USA}

\correspondingauthor{Alec Martin}
\email{anmg9n@mail.missouri.edu}

%\author{The UVCANDELS Team}



\begin{abstract}
    \indent Giant star-forming clumps are a prominent feature of star-forming galaxies (SFGs) and contain important clues on galaxy formation and evolution. However, basic demographics of clumps and their host galaxies remain uncertain. Using the HST/WFC3 F275W images from the Ultraviolet Imaging of the Cosmic Assembly Near-infrared Deep Extragalactic Legacy Survey (UVCANDELS), we detect and analyze giant star-forming clumps in galaxies at 0.5 $\leq$ z $\leq$ 1, connecting two epochs when clumps are common (at cosmic high-noon, z $\sim$ 2) and rare (in the local universe). We construct a clump sample whose rest-frame 1600 {$\AA$} luminosity is 3 times higher than the most luminous local HII regions (M$_{UV} \leq -$16 AB). In our sample, 35 $\pm$ 3$\%$ of low-mass galaxies (log[M$_{*}$/M$_{\odot}$] $<$ 10) are clumpy (i.e., containing at least one off-center clump). This fraction changes to 22 $\pm$ 3\% and 22 $\pm$ 4\% for intermediate (10 $\leq$ log[M$_{*}$/M$_{\odot}$] $\leq$ 10.5) and high-mass (log[M$_{*}$/M$_{\odot}$] $>$ 10.5) galaxies in agreement with previous studies. When compared to similar-mass non-clumpy SFGs, low- and intermediate-mass clumpy SFGs tend to have higher SFRs and bluer rest-frame U-V colors, while high-mass clumpy SFGs tend to be larger than non-clumpy SFGs. However, clumpy and non-clumpy SFGs have similar S\'ersic index, indicating a similar underlying density profile. Furthermore, we investigate how UV luminosity of star-forming regions correlates with the physical properties of host galaxies. On average, more luminous star-forming regions reside in more luminous, smaller, and/or higher-specific SFR galaxies and are found closer to their hosts’ galactic center.
\end{abstract}

\keywords{Galaxy evolution (594); Galaxy formation (595); Galaxy structure (622); Galaxy properties (615); Starburst galaxies (1570); Star forming regions (1565)}

\section{Introduction} \label{Intro}

    \indent Understanding galaxy formation and evolution requires tracking the evolution of different galactic sub-structures over cosmic time. One sub-structure of interest is giant star-forming clumps. In comparison to their host galaxies, these clumps have higher specific star-formation rate (sSFR) by a factor of several \citep{2012ApJ...757..120G,2012ApJ...753..114W,2017ApJ...837....6S}. Such clumps are common structures within star-forming galaxies (SFGs) at 1 $\leq$ z $\leq$ 2 that generally contribute, $\sim$10$\%$ of the total SFR within SFGs \citep{2012ApJ...757..120G,2015ApJ...800...39G}. Clumps are thus very prominent features  in rest-frame UV images and have been frequently detected through deep and high-resolution rest-frame UV and optical images with both space- \citep{2004ApJ...600L.139C,2007ApJ...658..763E,2009ApJ...701..306E,2014ApJ...786...15M,2017ApJ...837....6S} and ground- based telescopes \citep{2021ApJ...912...49M,2022ApJ...924....7S,2022ApJ...931...16A}. However, they have also been detected at other wavelengths, such as in emission-line maps of H$\alpha$ \citep{2008ApJ...687...59G,2011ApJ...733..101G,2016ApJ...831...78M,2017ApJ...839L...5F,2019MNRAS.489.2792Z} and as CO line emission detections from lensed galaxies \citep{2010MNRAS.404.1247J,2010MNRAS.405..234S,2017A&A...605A..81D}.

    \par
    \indent The physical properties of clumps have been studied in various ways. Previous work using multi-wavebands spectral energy distribution (SED) analyses with HST to study gravitationally lensed and unlensed clumps have suggested that they are much larger, brighter, and more massive than local star-forming regions. Several HST studies of unlensed galaxies have indicated that clumps have a typical mass range of log M$^{\ast}$/M$_{\odot} \simeq$ 8–9 \citep{2012ApJ...757..120G,2012ApJ...753..114W} and vary in effective sizes from $\sim$1 kpc \citep{2007ApJ...658..763E,2011ApJ...739...45F,2017ApJ...837....6S} to a few hundred pc \citep{2012MNRAS.427..688L,2017A&A...605A..81D}, while lensed studies have suggested much smaller clump masses of (log M$^{\ast}$/M$_{\odot} \simeq$ 6–7) and sizes ($\leq$100 pc) \citep{2017ApJ...843...79R,2018NatAs...2...76C,2021A&A...646A..57V,2022MNRAS.516.3532M}.
    \par 
    \indent Investigating this discrepancy, \citet{2018NatAs...2...76C} analysed multiple images of the same lensed galaxy at different magnification, finding that unlensed observations with $\sim$1 kpc resolution tend to overestimate both the size and luminosity of clumps, and therefore the mass due to the blending of smaller clumps into a single resolution element. Studies resolving this blending using local high-z SFG analogs of DYNAMO galaxies \citep{2017ApJ...839L...5F,2022MNRAS.512.3079A} and machine learning methods trained on high-resolution simulations and then applied to HST observations \citep{2020MNRAS.499..814H,2021MNRAS.501..730G} have found that giant clump masses are typically on the order of log M$^{\ast}$/M$_{\odot}$ $\simeq$ 7$-$8.

    \par
    \indent Along with the effect of the blending of clumps towards determining individual masses, clump research suffers from a lack of other demographic properties. This makes uncertain our understanding of clump formation/evolution and their overall contribution to their host galaxies development throughout their lifetimes. For example, the S\'ersic index of clumpy host galaxies is one of the key measurements for distinguishing the clump formation mechanisms. The two most widely accepted mechanisms are (1) in-situ formation from violent disk instabilities (VDI) and (2) ex-situ formation from major/minor merger events. The former predicts that clumps form in unstable regions where the Toomre Q parameter \citep{1964ApJ...139.1217T} is below an order of unity in gas-rich galaxy disks \citep{2007ApJ...670..237B,2009ApJ...707L...1B,2009ApJ...703..785D,2022MNRAS.511..316D,2009MNRAS.397L..64A,2010MNRAS.404.2151C,2014MNRAS.442.1230R,2015MNRAS.449.2156A}. Such galaxies are fed continuously by dense streams of cold gas flowing along cosmic web filaments. These remote large surface densities and high gas fractions lead to VDI, while also increasing the velocity dispersion which in turn increases the typical clump mass \citep{2009ApJ...703..785D,2021MNRAS.501..730G}. Alternatively, it is worth mentioning that for high redshift and Toomre Q parameters significantly above unity, compressive modes of turbulence can initiate clump formation through both in-situ and ex-situ scenarios, complicating our understanding of VDI at higher redshifts \citep{2016MNRAS.456.2052I}. However, a recent neural network detected clump study within CANDELS found matching clump stellar mass functions (cSMFs) for simulated and observed clumps at 1 $<$ z $<$ 3 \citep{2020MNRAS.499..814H}. This indicates that the majority of observed clumps within their sample formed through in-situ scenarios. Moreover, cosmological simulations of \citet{2014MNRAS.443.3675M,2017MNRAS.464..635M} suggests that ex-situ formation through merger events only accounts for at most $\sim$30\% of the overall clump population. Therefore, further examination of clump formation scenarios is needed.

    \par
    \indent For this purpose, characterizing the demographics can provide insight into clump formation. For example, if clumps are predominantly hosted by galaxies with S\'ersic index n $\approx$ 1, then this would be consistent with clump formation through VDI, which requires a disk morphology. Additionally, simulations from \citet{2014MNRAS.443.3675M,2017MNRAS.464..635M} show that a S\'ersic index n $\approx$ 1 places constraints on the stellar mass ratio of minor mergers as mergers above a ratio of 1:10 would very likely destroy the disk morphology. Furthermore, \citet{2017MNRAS.464..635M} show a far narrower distribution of SFR$_{c}$/SFR$_{g}$ for in-situ clumps than for ex-situ clumps, with the peak value for in-situ clumps around 10\%. This means that clumps are more likely to form through VDI if demographics show a correlation between the fractional (clump and host galaxy) SFR. Thus, further work is needed to characterize the demographics of clumps and their host galaxies.
    
    \par 
    \indent To obtain robust demographics of clumpy galaxies, large samples from deep surveys are necessary. An initial study of \citet{2014ApJ...786...15M} provided the clumpy fraction at 0.2 $\leq$ z $\leq$ 1 of optically selected clumpy and non-clumpy galaxies as a function of stellar mass and sSFR. \citet{2015ApJ...800...39G} provided their clumpy fraction for SFGs for a larger, UV-selected clump sample, but presented no demographics. Additionally, \citet{2020MNRAS.499..814H} studied a sample of clumps detected through a machine-learning algorithm that was tested on VELA simulated galaxies and utilized on SFG samples in the five CANDELS fields \citep{2011ApJS..197...35G,2011ApJS..197...36K}. This study contributed their clumpy fraction of SFGs in terms of host mass, $\Delta$log(sSFR), $\Delta$log(R$_{e}$) along with correlations between clump properties and their host galaxies, but not for a wide range of demographics. \citet{2016ApJ...821...72S} directly compared clumpy and non-clumpy galaxies that were UV and optically selected for both SFGs and quiescent galaxies (QGs). Their main demographic results show that clumpy galaxies are statistically different from non-clumpy galaxies at 1 $\leq$ z $\leq$ 2 in terms of host galaxy morphology for all stellar masses, and SFR for low- and intermediate-mass galaxies.
    \par 
    \indent Within this work we utilize the mosaics from a recent UV survey, UVCANDELS (PI: H. Teplitz), that has been conducted on 4 of the 5 CANDELS fields (COSMOS, GOODS-N, GOODS-S, EGS). UVCANDELS, when combined with CANDELS, includes full coverage from NUV ($\sim$2600 \AA) to I-band ($\sim$8000 \AA) filters providing ideal mosaics to study UV-bright clumps in the lower redshift regime (z $\leq$ 1). We detect and analyze a sample of UV-selected clumps from these mosaics to further investigate the demographics of clumps and characteristic differences between clumpy and non-clumpy SFGs. By defining a clumpy galaxy as an SFG that contains at least one off-center clump, we are able to separate our UV-selected sample into clumpy and non-clumpy populations and analyze a large variety of averaged galactic properties (Host: stellar mass; optical size; UV color; SFR; and S\'ersic index) relations as discussed further in Section \ref{Results}. We compare these results directly to those of \citet{2016ApJ...821...72S} and discuss the implications of any differences. Furthermore, we present several relations between clump luminosity and their host galaxy properties to further relate clumps and the types of galaxies that exhibit them. We also note that previous studies are not homogeneous in their definition (or constraints) of clumps, which could play a role in the demographic discrepancies present within the literature. To this end, our results are tested across several definitions for clumps and SFGs to show the impact of sample selection. We compare our findings to previous studies in Section \ref{ESS}.

     %%PLACED HERE FOR COSMETICS
    \begin{figure*}[htb!]
        \centering
        \includegraphics[scale=0.7]{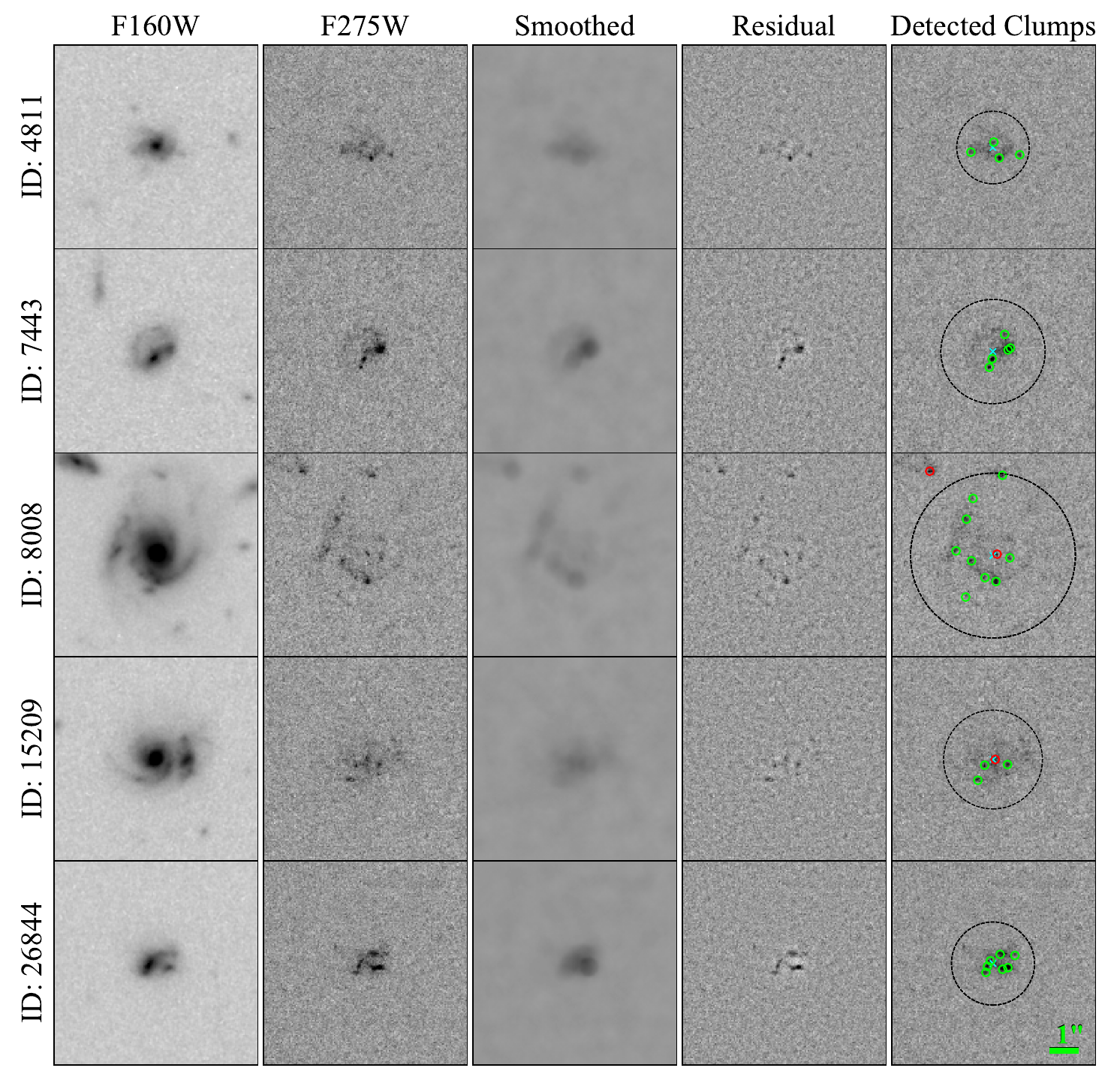}
        \caption{Example galaxies (rows) from our clump detection algorithm. The respective columns from left-to-right show each galaxy in the H-band filter, Near-UV filter, Smoothed (F275W), Residual (F275W - Smoothed), and original F275W cutout with detected clumps marked. Each galaxy center is marked with a cyan `x' and our clump detection size limit of r $\leq$ 3$\times$R$_{e}$ indicated by black dashed circles. The detected clumps that pass(fail) our fiducial selection criterion are marked with green(red) circles. Our selection criterion removed 1 clump from both ID: 8008 and 15209 due to their proximity to the host galaxy's center. Additionally, another clump was removed from ID: 8008 due to its radial distance being greater than 3 R$_{e}$, which suggests a probable association with an external galaxy.}
        \label{det}
    \end{figure*}

    \par
    \indent We structure this paper as follows. First, we briefly summarize our surveys and SFG samples in Section \ref{Data}. We discuss the method for clump detection within individual galaxies from their UV mosaic cutouts in Section \ref{CD}. In Section \ref{photometry}, we describe the measurement of multi-band photometry of individual clumps and test its accuracy. We then present our work's contribution to the clumpy fraction across cosmic time in Section \ref{cf}. In Section \ref{cvnc} we present our demographic results between clumpy and non-clumpy SFGs and compare to the literature. In Section \ref{ESS}, we discuss the effect of sample selection and our definitions of clumpy and quiescent galaxies that potentially impose biases. In Section \ref{cfvgp} we present our clump luminosity relations and compare these results to the literature. In Section \ref{Discussion}, we discuss the implication of our results towards the evolution of SFGs that contain clumps while noting some contaminating scenarios not studied by this work. Lastly, we conclude this paper while describing our future work. Throughout this paper we adopt a flat $\Lambda$CDM cosmology with $H_{0}$ = 67 km/s/Mpc, $\Omega_{M}$ = 0.31, $\Omega_{\Lambda}$ = 0.69 and express all magnitudes in the AB system \citep{1974ApJS...27...21O}.
        
\section{Data and Methods} \label{Data}
    \subsection{Surveys \& Galaxy Properties}

    \par
    \indent The samples of galaxies used in this work are derived from the mosaics of the recently completed Hubble Treasury program, the Ultraviolet Imaging of the Cosmic Assembly Near-infrared Deep Extragalactic Legacy Survey Fields (UVCANDELS). This survey is an extension of CANDELS \citep{2011ApJS..197...35G,2011ApJS..197...36K} in UV coverage for four of the deep-wide survey fields (COSMOS, GOODS-N, GOODS-S, and EGS). UVCANDELS consists of WFC3/F275W imaging (imaging depth of AB = 27 mag for compact galaxies) over $\sim$430 arcmin$^{2}$ of UV coverage, an increase of 4x when combined with prior archival data, along with ACS/F435W (reaching AB $\geq$ 28.0 mag) imaging in parallel.
    
     \par 
    \indent As a comparison sample we use galaxies selected from the F275W mosaics of the Hubble Deep UV Legacy Survey (HDUV) which is a deeper survey ($\sim$0.5 mag deeper in comparison to UVCANDELS at z $\sim$ 1) covering the GOODS-N and GOODS-S CANDELS fields. The additional depth allows us to test for survey sensitivity bias within the clump detection process. We refer the reader to \citet{2018ApJS..237...12O} for a more thorough description of the HDUV survey.
    
    \par
    \indent For our analysis we utilize the galactic properties listed in the original CANDELS catalogs \citep{2011ApJS..197...35G,2011ApJS..197...36K} matched to our galaxies from both UVCANDELS and HDUV. In brief, the derived photometric redshifts combine the results from more than a dozen photo-z measurements with various SED-fitting codes and templates, which is further described in \citet{2013ApJ...775...93D}. Stellar masses were measured as the median of the results from 12 SED-fitting codes, as adopted from \citet{2015ApJ...801...97S} for GOODS-S, \citet{2017ApJS..229...32S} for EGS, \citet{2017ApJS..228....7N} for COSMOS and \citet{2019ApJS..243...22B} for GOODS-N. The SFRs were derived from UV luminosities using the relation from \citet{1998ApJ...498..541K} and corrected for dust using the slope of the UV continuum emission to the ratio of UV to IR luminosities as described in \citet{2019ApJS..243...22B}. Galaxy size and morphology were derived using GALFIT with the H, J, and Y-bands detailed in \citet{2012ApJS..203...24V}. Lastly, we determined color for our UV-bright SFGs by using the CANDELS cataloged rest-frame U- and V-band AB magnitudes, derived from observed-frame spectral energy distributions (SEDs) template fitting using the EAZY code \citep{2008ApJ...686.1503B}.

    \subsection{Sample Selection}
           
        \indent We first constrained our UVCANDELS and HDUV galaxy samples by their listed CANDELS physical properties. A redshift range of 0.5 $\leq$ z $\leq$ 1 was selected in order to avoid observing the UV bright clumps at optical wavelengths and when clumps are rare ($z < 0.5$). A minimum galaxy mass limit was set to log(M*) $\geq$ 9.5 $M_{\odot}$ to ensure a mass-complete sample. An axial ratio of q $>$ 0.5 excludes ``edge-on'' galaxies minimizing the effects of dust extinction in our flux calculations in Section \ref{photometry}. To avoid contamination by stars we set Star Class $<$ 0.8, removing likely star candidates above 80\% probability. We restrict our sample to actively star-forming galaxies by defining SFGs as galaxies with sSFR $\geq$ 0.1 Gyr$^{-1}$. Lastly, we placed constraints on galaxy size (R$_{e}$ $\geq$ 0.$\arcsec$2) and H-band magnitude (F160W $<$ 25 AB) to remove small and faint galaxies, as their sub-structures may not be properly resolved. The above restrictions reduced our final sample to 513 SFGs for the 4 fields in UVCANDELS and 182 SFGs for the 2 fields of HDUV with 4 overlapping galaxies.
            
\subsection{Clump Detection} \label{CD}

    \indent  Clumps were detected from the HST/UVIS F275W 60-mas mosaics through an automated detection algorithm, with quality control by visual inspection. Within the four UVCANDELS fields and two HDUV fields, we made a 150$\times$150 pixel cutout image of each galaxy and convolved them with a Tophat 2D-smoothing kernel set to r=7 pixels (0.$\arcsec$42). From the resulting residual image (original $-$ smoothed), the median global background contribution was measured with a 25$\times$25 pixel box and all pixels with a flux value less than 3$\sigma$ masked out, leaving only UV-bright pixels that are potentially associated with clumps. From this, we identified sources that contain a FWHM of 3 pixels brighter than 3$\sigma$ and labeled them as potential clumps. In order to verify the robustness of our algorithm, we experimented with a wide range of values for the roundness, threshold background-subtracted flux value, and minimal FWHM for clumps. Visual inspection of each galaxy for each set of such parameters confirms that our algorithm consistently detects bright clumps in both UVCANDELS and HDUV. Figure \ref{det} shows the detection algorithm process for 5 example galaxies and includes their corresponding H-band image for reference.
    
    \par
    \indent Within our samples not all of the detected clumps are of interest. Specifically, we excluded: (1) central clumps; and (2) clumps not associated with the target galaxy. The former exclusion removed galactic nuclei misidentified as clumps by flagging the detected sources with radial distance $\leq$ 0.$\arcsec$1 from the CANDELS listed galactic center. The latter exclusion removes clumps or other bright sources that originated within other non-targeted galaxies/objects also present within a cutout. This restriction is set by mandating a clump be within a radial distance of 3 times the CANDELS listed effective radius (R$_{e}$) in arcseconds. Using similar methods to that of \citet{2005PASA...22..118G}, we find that a radius of r = 3$\times$R$_{e}$ for n=1(4) encompasses 98.2\%(93.4\%) of the galaxy’s total flux allowing for detections across the majority of the target galaxy regardless of their morphology. Visual inspection of both UVCANDELS and HDUV samples were again performed to validate both distance restrictions.
    %and additionally remove any clumps from regions within obvious mergers.
    
    \par
    \indent Additionally, we investigated the possibility that our clump statistics are contaminated by the superimposition of foreground/background galaxies within our mosaics. To accomplish this, we used the number density of observed galaxies in UVCANDELS in each sky area to calculate the probability of two galaxies falling into the same area as large as a typical galaxy in our sample. More specifically, to estimate the number density, we included all UVCANDELS galaxies at 0.5 $\leq$ z $\leq$ 2.0 with F275W magnitude brighter than 28 AB. We limited the redshift to z=2.0, because beyond it the Lyman limit will be redshifted into the F275W filter so that the number of observed galaxies would be quite low (see Wang et al., in prep). We also used sources brighter than 28 AB because our clump detection completeness drops to nearly zero for fainter clumps according to our test involving simulated clumps in Section \ref{photometry}. We determined that the average number of galaxies in a circular region with radius of 3$\arcsec$ is $\sim$0.1. This radius is 3 times the maximum radius of all galaxies in our sample (1$\arcsec$). Assuming a Poisson distribution, with this average number, the probability of observing two galaxies simultaneously in such a region is 0.45\%, meaning only $\sim$2 of our 513 SFGs would be affected by the foreground/background contaminations. This result is negligible for our later statistical analysis. Visual inspection of each UV mosaic also indicated no obvious merger pairs within our sample.

\subsection{Aperture Photometry} \label{photometry}
    \par
    \indent For our clump photometry we measure the flux within an aperture of radius 0.$\arcsec$18 (3 pixels) for each clump. Since clumps are not isolated sources, other clumps can fall within each defined aperture/annulus. Therefore, while measuring the flux for individual clumps, all others are masked out from our calculations.
 
    \par
    \indent Once the total aperture flux is measured we apply two corrections. First, since clumps are sub-structural features that lie embedded within the gaseous disks of galaxies, the local background contribution (measured using an annulus of size 0.$\arcsec$3 $-$ 0.$\arcsec$42 [5$-$7 pixels] with clumps masked from the image) was subtracted from the total to exclude the flux contributed from the disk profile. Both our aperture and annulus size are slightly larger than the fiducial method in \citet{2018ApJ...853..108G}, but still provide reasonable estimates. We acknowledge that our chosen 3-pixel-sized masks do not fully block the signal contributions from the PSF wings of other clumps within our defined background annulus. However, at a radius of r$>$3 pixels, the PSF flux per pixel is less than 2\% of its central maximum flux. Here, for simplicity, we assume each clump is a PSF. Moreover, since a nearby clump (if present) only occupies one direction of the entire background annulus, our use of median to estimate the background with the entire annulus would not be significantly affected. Therefore, the contribution from the PSF wings of nearby clumps is negligible in our later measurement of clump fluxes.
    
    \par
    \indent Secondly, our choice of aperture size might not encapsulate the entire point spread function (PSF) of our sources, and therefore may miss the contributions from the PSF-wings of their light profiles. To correct for the light outside of our defined apertures, we estimated a correction factor by calculating the ratio of flux encircled within our 3-pixel aperture to the total flux of the PSF. We determined that our aperture encapsulates $\sim$77\% of the total light from each source leading to a global correction factor of 1.29 for all four CANDELS fields.
    
    \begin{figure}[htb!]
        \centering
        \includegraphics[scale=0.53]{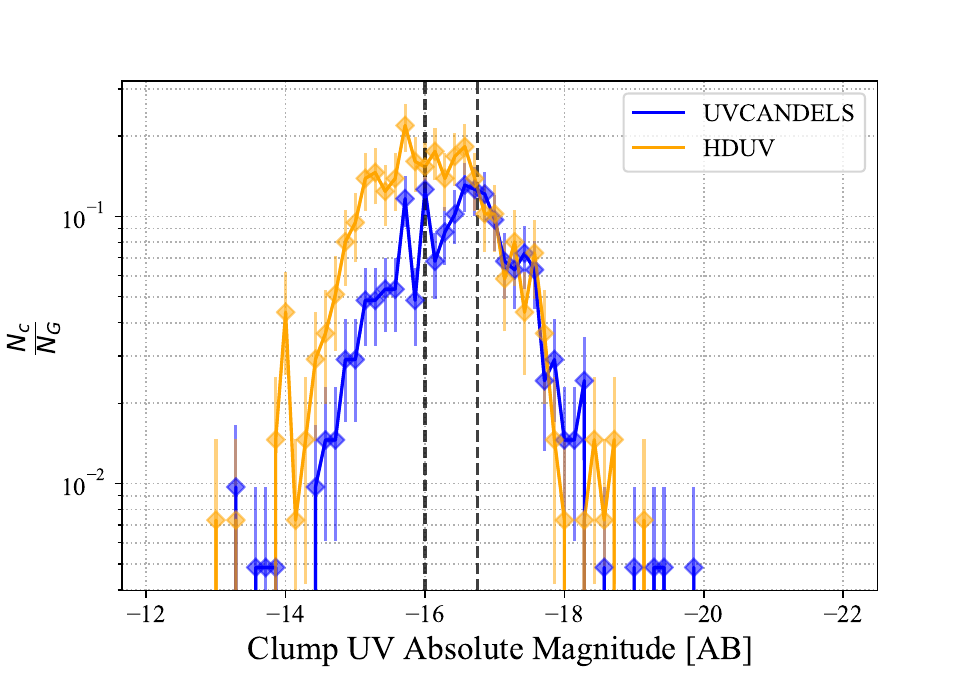}
        \caption{Clump luminosity comparison of UVCANDELS (blue line) to HDUV (orange line) with their corresponding 1$\sigma$ Poisson error. The black dashed lines represent two sample cuts at clump UV luminosity (expressed in absolute magnitude) M$_{UV}$ = $-$16 and $-$16.7 AB. The former represents a sample distinct from typical HII regions that was used as our fiducial results. The latter represents a complete sample in which detections for UVCANDELS is not affected by lower sensitivity limits.}
        \label{flim}
    \end{figure}

    \par
    \indent To evaluate the completeness of our clump detection, we compare in Figure \ref{flim} the ratio of the number of clumps to the number of SFGs (N$_{c}$/N$_{g}$) in UVCANDELS and HDUV, with the latter being about $\sim$0.5 mag deeper than the former. Clumps from HDUV have the same method of detection and luminosity measurement as those from UVCANDELS. We converted the flux of each clump to the absolute magnitude in rest-frame UV ($\lambda \sim$1600\AA) following the method of \citet{2002astro.ph.10394H}. For simplicity, we assumed that clumps have the same stellar population and dust as their host galaxies, simplifying the conversion process to only require the host galaxy's redshift. This assumption is, in general, valid for clumps as a population (e.g., \cite{2012ApJ...757..120G,2018ApJ...853..108G}). Although individual clumps may have different M$_{*}$/L ratios (e.g., varying with their galactocentric distances), but this assumption would not affect our completeness estimate here. The two surveys match at the bright end down to $\sim -$16.7 Mag (right-most dashed line), indicating that the shallower UVCANDELS still yields a complete detection of clumps above this luminosity. For luminosities fainter than $-$16.7 Mag, the number of clumps per galaxy in UVCANDELS is significantly lower than in HDUV, indicating an incomplete clump detection.
    \par 
    \indent To ensure complete statistics in our later analyses, we set a minimum clump luminosity limit derived from a comparison to the far-UV luminosity of the largest local HII regions from \citet{2009ApJ...699.1125R}. More specifically, we derive the far-UV luminosity of NGC 604 from \citet{2009ApJ...699.1125R} from their extinction-corrected H$\alpha$ luminosity with the assumption that SFR$_{H\alpha} \approx$ SFR$_{FUV}$. We find that clumps with luminosity M$_{UV} \leq -$16 AB (left-most dashed line of Figure \ref{flim}) are $\sim$3 times brighter than the most luminous local HII regions (in comparison to NGC 604), ensuring that our clumps are intrinsically different from typical HII regions. This limit provides the highest number statistics for our clump sample, while also maintaining a reasonable completeness limit for our detections ($\sim$60\% at M$_{UV}$ = $-$16 AB) as shown in Figure \ref{flim}. We set clumps brighter than this threshold as our fiducial sample, however, we note that this sample remains incomplete when compared to the HDUV detections. We will also discuss further in Section \ref{ESS} the effects of using a complete sample (M$_{UV}$ $\leq -$16.7 AB limit) that suffers from lower number statistics and compare the results to our fiducial sample. Additionally, we compare other sample selection criteria to our results, for completeness.
    
    \par
    \indent Furthermore, we have conducted a thorough test of our clump detection algorithm to evaluate its completeness, by using a method similar to \citet{2015ApJ...800...39G} and \citet{2023AAS...24124907S}. Our test involved the use of simulated clumps, which were represented as point-like sources with normalized magnitudes between F275W mag = 25$-$30 AB. These clumps were embedded within the cutout mosaic of a specific non-clumpy, HDUV galaxy that is characterized by a disk profile. We then tested our detection algorithm, processing the new image in a similar manner as described before, and repeated this approach 100 times for each incremental step in clump magnitude for the full redshift range of 0.5 $\leq$ z $\leq$ 1. For comparison to our fiducial cutoff of M$_{UV} \leq -$16 AB, we converted the flux of our simulated clumps to absolute magnitudes using the same method as before for our observed clumps. Our findings indicate that our detection algorithm achieves $\sim$100\% completeness for clumps down to our fiducial cutoff at the median redshift of z $\sim$ 0.8 for our HDUV sample. We therefore claim that our clump detection algorithm is complete within our study.

    \begin{figure}[htb!]
        \centering
        \includegraphics[scale = 0.6]{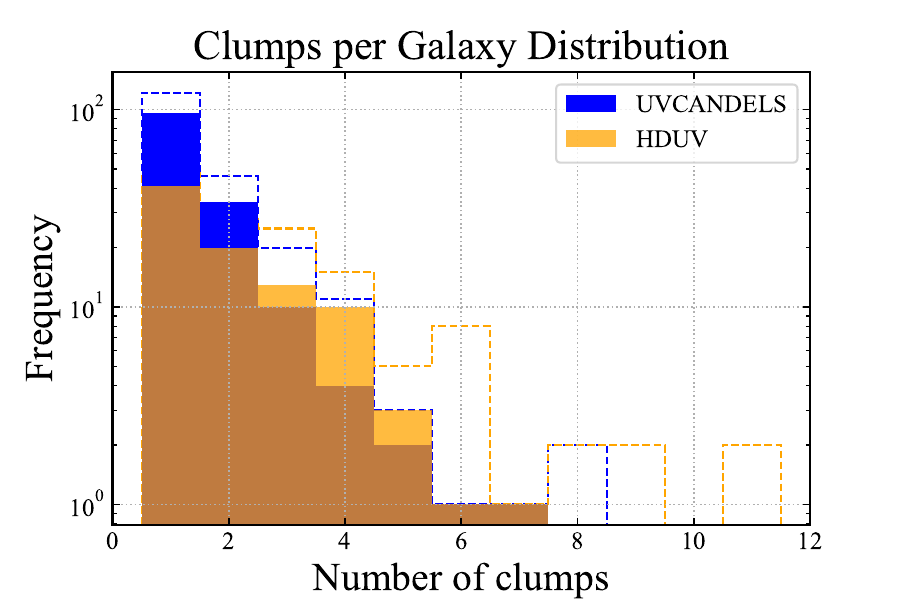}
        \caption{Frequency of detected clumps per galaxy for our UVCANDELS (blue) and HDUV (orange) SFG samples. The dashed bars represent all detections, while the solid bars represent those clumps that pass the selection criterion for our fiducial sample.}
        %our more stringent selection criterion.}
        \label{Cnum}
    \end{figure}

    %%CLUMPY FRACTION PLACED HERE FOR COSMETICS%%
    \begin{figure*}[htb!]
        \centering
        \includegraphics[scale=0.6]{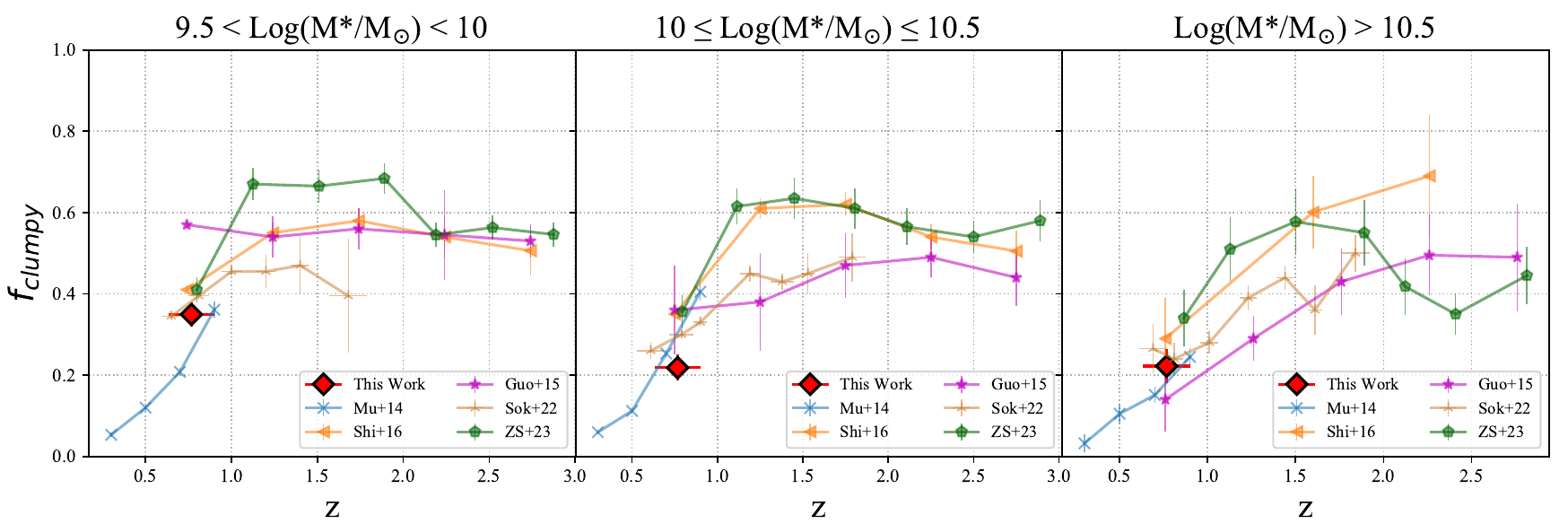}
        \caption{Clumpy Fraction for [left-to-right] low/intermediate/high mass galaxies. Our UVCANDELS results (outlined red diamonds) show good agreement with the UV selected samples (colored lines) for all mass bins.}
        \label{CF}
    \end{figure*}
    
    \par
    \indent In Figure \ref{Cnum} we show the number of clumps per galaxy for our two survey samples that were detected from our algorithm. After applying our selection criterion our fiducial sample (filled bars) consists of 233 and 188 clumps for UVCANDELS and HDUV respectively with the majority of SFGs in UVCANDELS (88\%) and HDUV (69\%) hosting 1$-$2 clumps. The most clumpy galaxy seen within each survey both host 7 clumps within each of their substructures.

\section{Results} \label{Results}
    
    \par
    \indent In this section, we present our main results: clumpy fraction, properties of clumpy galaxies, and relations between clump luminosity and host galaxies’ properties. We then discuss the effects of different sample selections and clump definitions on our results. To investigate clump properties in different galaxy M$_{*}$ regimes, we divide our sample into three mass bins (low mass: 9.5 $<$ M$_{*}$/M$_{\odot}$ $<$ 10; intermediate mass: 10 $\leq$ M$_{*}$/M$_{\odot}$ $\leq$ 10.5; and high mass:  M$_{*}$/M$_{\odot}$ $>$ 10.5) throughout our work. 
    
    \subsection{Clumpy Fraction}\label{cf}
        \indent To evaluate the importance of clumpy sub-structures to the evolution of SFGs, we measure the fraction of clumpy galaxies, hereafter ``f$_{clumpy}$'', defined as the number of SFGs containing at least one off-center clump, to the total number of SFGs in our sample. Figure \ref{CF} shows our result (outlined circles) and their corresponding binomial errors within our redshift range. For low mass galaxies (left panel) $\sim$ 35\% of SFGs are considered `clumpy' while the fraction decreases to $\sim$ 22\% for both intermediate (middle panel) and high mass (right panel) galaxies. As indicated, the frequency of clumpy SFGs decreases between cosmic noon and the present day. However, a significant fraction of SFGs are clumpy at z $\geq$ 0.5 regardless of stellar mass. 
         \par 
         \indent We compare our results to those within the literature in Figure \ref{CF}. \citet{2014ApJ...786...15M} (blue lines) derived f$_{clumpy}$ from optically (HST/F814W) detected galaxies in the COSMOS field. Their sample of galaxies covered a similar redshift range to ours, however, they defined their clumpy galaxies differently based on the I-band flux ratio of the 3 brightest clumps in each galaxy. \citet{2015ApJ...800...39G} sampled galaxies from the CANDELS/GOODS-South and UDS fields, while \citet{2016ApJ...821...72S} sampled galaxies from all five CANDELS fields, along with data in the Hubble Ultra Deep Field (HUDF) \citep{2006AJ....132.1729B} and 2 parallel Hubble Frontier fields (HFF) Abell 2744 and MACS0416 \citep{2017ApJ...837...97L}. Both studies used a similar sample selection, with the exception of slightly different binning for low/intermediate/high stellar mass, and defined their clumps using the ratio of clump UV luminosity to the total UV luminosity of the host galaxy (e.g. $L_{c}/{L_{G}} \geq$ 8\%). \citet{2023AAS...24124907S} studied clumps within UVCANDELS and CANDELS that contribute a minimum of 10\% of the galaxy’s total rest-frame UV flux and are detected by their Fast Fourier Transform (FFT) based algorithm. Their results are consistent with other rest-frame UV detected samples. Lastly, \citet{2022ApJ...924....7S} used deconvolution of ground-based images of the COSMOS field to resolve 20,185 SFGs at 0.5 $\leq$ z $\leq$ 2. They detected sources as high, U-band surface density regions within the plane of surface brightness density and host galaxy size ($\Sigma$/$\Sigma_{e}$ , R/R$_{e}$ ) and similarly defined them as `clumps' if their fractional U$_{rest}$ luminosity was above 8\%. We find that our measured clumpy fraction agrees very well with the literature for each mass bin.

         \par 
         \indent We note the importance of detecting clumps in the UV as detections at redder wavelengths tend to under-sample clumps due to their lower contrast with their host galaxy's disk profile \citep{2012ApJ...753..114W}. This has been verified through morphology studies of Concentration, Asymmetry, and Smoothness (CAS) parameters from \citet{2018ApJ...864..123M} in which concluded a mild decrease in clumpiness (smoothness) within a wide range of galaxy morphologies when switching observations from UV to optical wavelengths. This effect is best shown from our results in the low mass bin of Figure \ref{CF}, for which the optical selected (I-band) clumps of \citet{2014ApJ...786...15M} reported a lower clumpy fraction.
        
    \subsection{Clumpy vs Non-Clumpy: Galactic Property - Stellar Mass Relations}\label{cvnc}
            
        \begin{figure}[htb!]
            \centering
            \includegraphics[scale = 0.63]{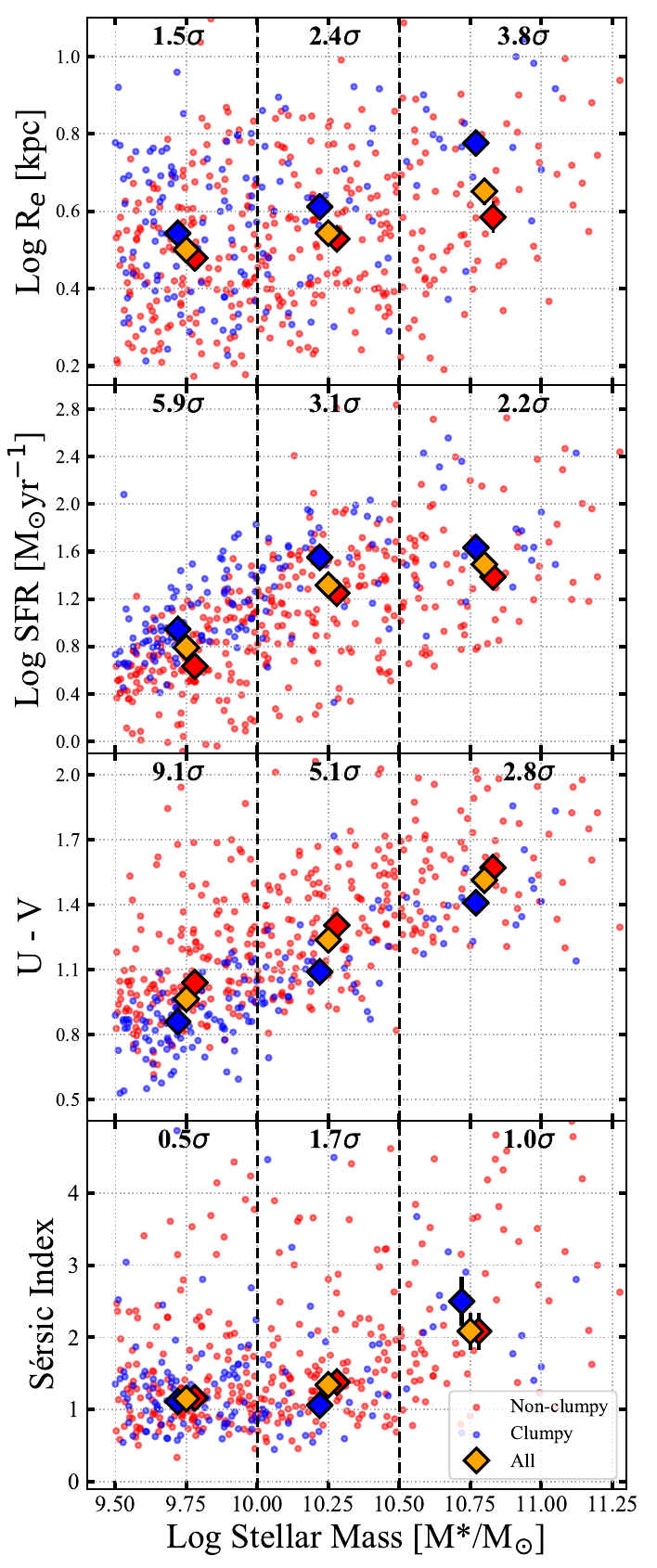}
            \caption{Galaxy properties $-$ stellar mass relations for (top to bottom) size, SFR, color and morphology. The blue/red points in each panel represent clumpy/non-clumpy SFGs with their median values given by the diamonds of the same color. The orange diamonds indicate the median of the two combined populations. The corresponding error bars are shown although most are smaller than the median symbol. The vertical black lines represent mass bin boundaries for low, intermediate and high mass SFGs as defined in the text. The statistical difference between the respective medians is listed at the top of each bin.}
            \label{results}
        \end{figure}
            
        \indent In Figure \ref{results} we plot several galactic properties (host size, SFR, color and S\'ersic index) against galaxy stellar mass for our clumpy and non-clumpy SFG distributions. The median of each population in each mass bin was calculated and are represented by diamonds within the figure. The uncertainty of each median was determined from bootstrapping with replacement over N = 1000 trials for both clumpy and non-clumpy populations. In most cases, the error bars are smaller than the median symbols (diamonds). To quantify the difference between clumpy and non-clumpy SFGs, we calculated the statistical significance between the two distributions (shown at the top of each panel) which is the difference in medians normalized by the quadrature sum of their respective uncertainty.
        
        \par 
        \indent We find that the difference between the clumpy and non-clumpy galaxies are statistically significant ($\geq$3$\sigma$) for host galaxy size in high mass SFGs (top panel), host log(SFR) (second panel) and color (third panel) in low and intermediate mass SFGs. The bottom panel of Figure \ref{results} indicates no significant differences between the S\'ersic index of clumpy and non-clumpy SFGs, with clumpy SFGs of low and intermediate mass showing clear disk-like structures (n $\approx$ 1). We argue that the slightly higher S\'ersic index (n $\sim$ 2.5) seen for high mass SFGs results from massive galaxies containing a more developed central bulge than lower mass galaxies, however still represents a disk-like morphology.
            
        \begin{figure*}[htb!]
            \centering
            \includegraphics[scale =0.74]{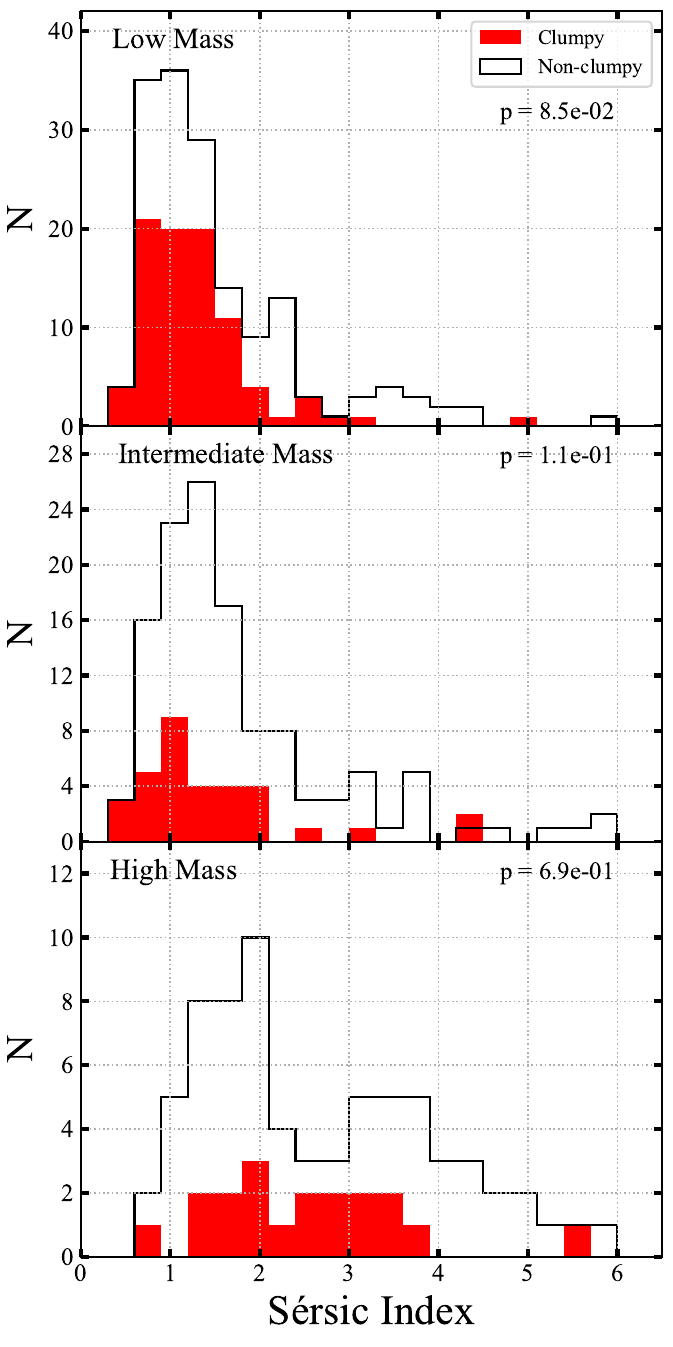}
            \qquad
            \includegraphics[scale=0.74]{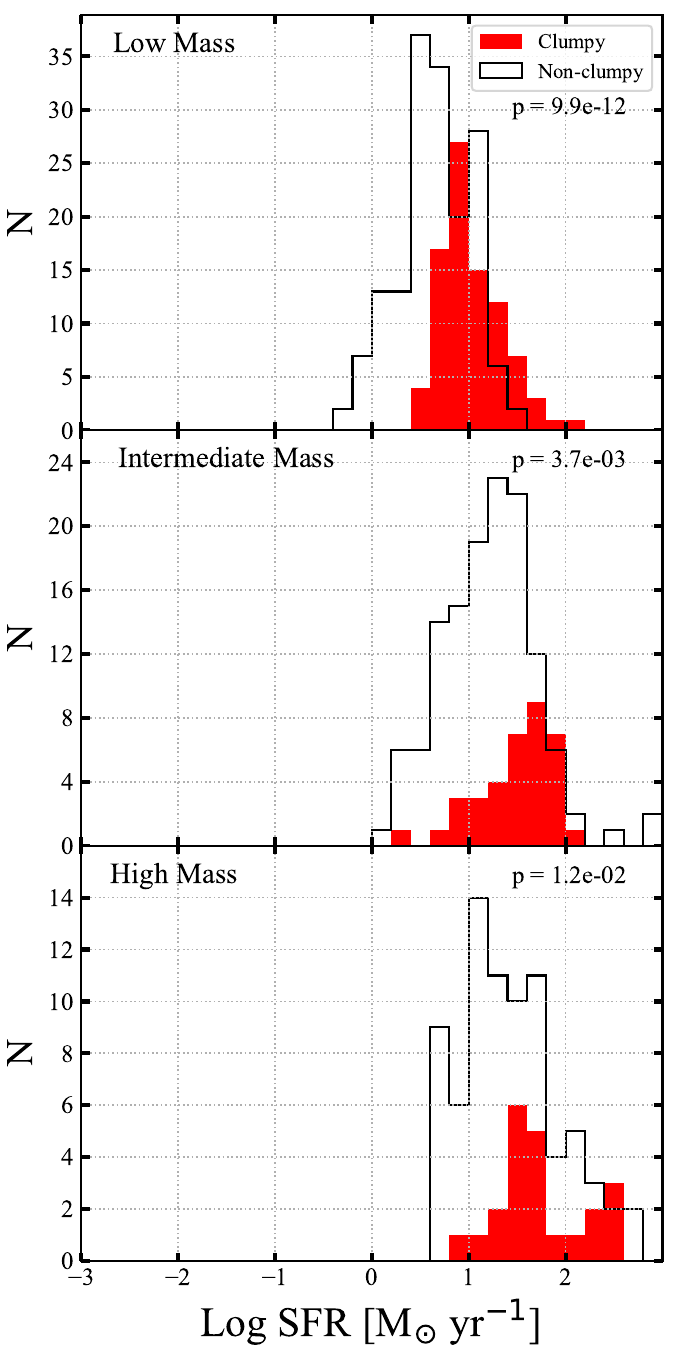}
            \caption{Histograms for S\'ersic index and SFR to be compared to \citet{2016ApJ...821...72S}. The figure shows (top to bottom) the low, intermediate, and high mass bins for S\'ersic index (left) and SFR (right). The red(outline) histograms represent clumpy(non-clumpy) distributions respectively, and the Two-Sample Kolmogorov-Smirnov Test p-value is shown within each panel.}
            \label{lit_comp}
        \end{figure*}
            
        \indent To compare to the literature, we further placed our results for SFR and S\'ersic index into histograms as shown in Figure \ref{lit_comp}. We display our clumpy (red) distribution and non-clumpy (outlined) distribution from top-to-bottom for low mass, intermediate mass and high mass galaxies respectively. We also calculate the two-sample K-S parameter in each panel indicating whether the two distributions for clumpy and non-clumpy SFGs are statistically similar (p $>$ 5\%) or distinct (p $\leq$ 5\%). We find consistent results for host morphology to our median comparison in that clumpy galaxies are statistically similar in disk-like morphology. Conversely, the distributions of clumpy and non-clumpy SFGs are distinct for host SFR in all mass bins.
        
        \par 
        \indent We compare these results with those found within \citet{2016ApJ...821...72S}, who used similar criteria to select galaxies from CANDELS, HUDF, and HFF as mentioned in Section \ref{CF}. Their results for SFR and S\'ersic index show a distinction for galaxies in all mass bins, with the exception of high mass SFR. Conversely to theirs, our clumpy and non-clumpy distributions are distinct for SFR and identical for morphology in all mass bins. However, their study includes quiescent galaxies (QGs) which have a higher S\'ersic index and are predominately non-clumpy. Therefore, including them will increase the average S\'ersic index of non-clumpy galaxies, resulting in the apparent distinction between the two populations. We note the existence of a second `peak' in the distribution of host-galaxy SFRs for high-mass clumpy galaxies. This may be due to poor statistics (22 galaxies) for clumpy galaxies in this mass bin. Therefore, it is possible that insufficient statistics result in the discrepancy in SFR between our results for high mass SFGs.
        
        %% SURFACE BRIGHTNESS PLACED HERE FOR COSMETICS %%
        \begin{figure*}[htb!]
            \centering
            \includegraphics[scale=0.69]{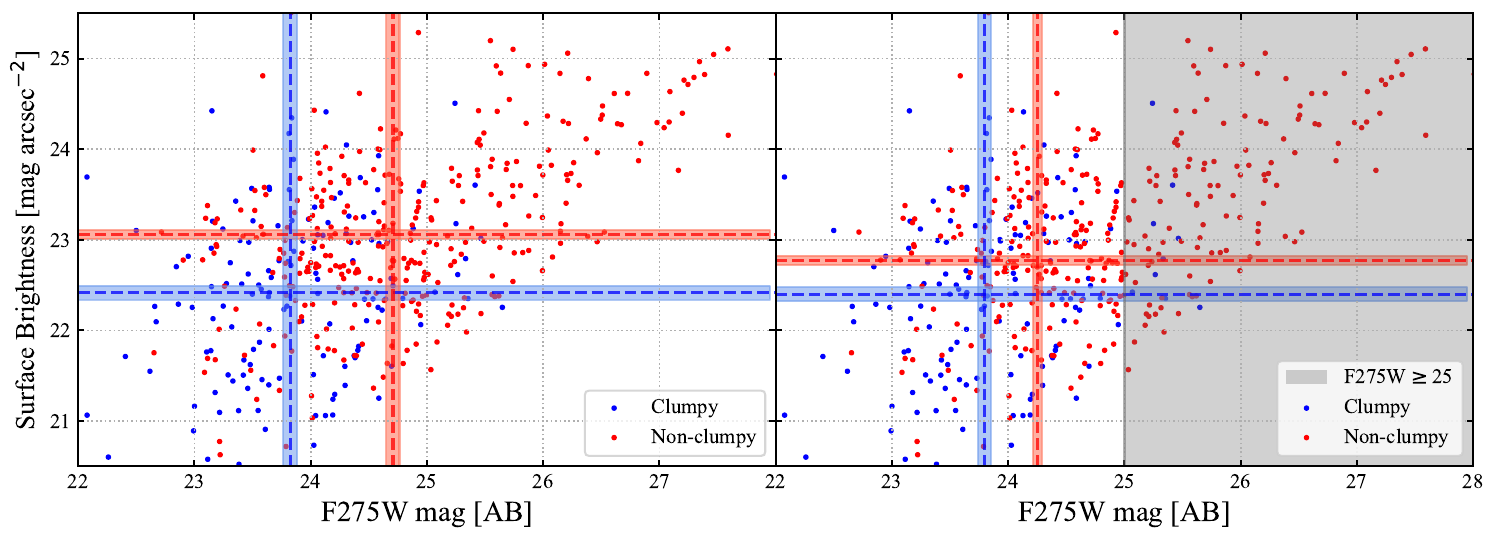}
            \caption{Fiducial surface brightness distribution before our F275W magnitude restriction (left panel) and after (right panel). The blue/red points represent clumpy/non-clumpy galaxies with the dashed lines and corresponding shaded regions representing their respective medians and errors. The right panel shows the same distribution as the left with the medians now representing the sub-sample of data points with F275W mag $\leq$ 25. The grey shaded region represents the data points that were excluded in the new median calculations. With the magnitude restriction the difference between clumpy and non-clumpy surface brightness is below 0.3 dex indicating a sufficient level of sample completeness.}
            \label{SB}
        \end{figure*}    
            
    \subsection{Effect of Sample Selection and Clump Definitions}\label{ESS}
        
        \indent In this section we discuss several effects of sample selection and clump definition on our results for our fiducial (M$_{UV} \leq -$16 AB) sample. Previous studies of clumps have defined their galaxies/clumps  using different methods. Furthermore, how we define `clumps' and `SFGs' can be important factors when comparing past and future studies and the impact of using different methods has not been fully explored. The purpose of this section is to provide a reference of these effects to make direct and fair comparisons in the literature, where the selections of clumps or clumpy galaxies vary from work to work. We provide our full range of sample selections in a single plot (Figure \ref{sigma_SS}) in the appendix for further reference.
            
        \subsubsection{UVJ vs. sSFR Quiescent Cutoff}
            \par
            \indent Here we specifically test two quiescent cutoff definitions for SFGs and quantify the impact to each of our two mass binned distributions. The first method uses a sSFR limit (0.1 Gyr$^{-1}$) that is sufficiently lower than the SFR-M* main-sequence, while the second method uses a color-color relation referred to as the ``UVJ-Diagram" \citep{2009ApJ...691.1879W}.
            
            \par
            \indent Compared to our fiducial sample, the UVJ definition for SFGs reclassifies (2, 14, and 16) SFGs as QGs in our low, intermediate and high mass bins respectively. Conversely, (6, 8, and 14) QGs would then be considered star-forming. Therefore, out of the total number of galaxies (SF $+$ QG), 3\%, 10\% and 16\% of each respective mass bin were defined differently as either QGs or SFGs between the two methods. Figure \ref{sigma_SS} located in the appendix shows this impact to our results more clearly. In comparison to our fiducial sample, we find that the UVJ definition only affects our conclusions regarding host-galaxy color for high-mass SFGs while all other results for size, SFR, color and morphology are unaffected. Therefore, we conclude that the particular quiescent galaxy cut-off definition used within a study has a marginal effect on SFG demographic results. 
            
        \subsubsection{Surface Brightness Bias} \label{sb}
        
            \indent One concern is that the difference in demographics may be affected by a difference in surface brightness for our SFG samples. For example, if clumpy galaxies have higher surface brightness on average this may be an indication that clumps are only detected above a certain brightness threshold. One possible solution to this problem would be to implement a minimum F275W magnitude to our SFGs so that our clumpy and non-clumpy samples are relatively similar with respect to their range in surface brightness distributions.
                
            \par
            \indent Figure \ref{SB} shows the surface brightness to apparent UV magnitude relation for our fiducial sample distribution with the medians and errors of our clumpy (blue) and non-clumpy (red) samples superimposed. We tested several UV-magnitude cutoff limits and set a range of F275W mag $\leq$ 25 as the optimal limit as shown by the grey shaded region in the right panel with the recalculated medians. This limit lowered the discrepancies between the median surface brightness of the two populations to below 0.3 dex indicating that the bias has been significantly reduced.
            %This limit reduces the difference in surface brightness between the two populations to below 0.3 dex indicating that the bias has been removed.
            
            \par
            \indent The impact of this magnitude constraint on our results can be seen in the left column of Figure \ref{results_25_8p} located in the appendix. The panels again show the relations of galaxy size, SFR, color and morphology with respect to galaxy stellar mass for our two populations. We find that the distinction between clumpy and non-clumpy SFGs for intermediate mass SFR and color as well as high mass size are no longer present. Furthermore, no new distinctions emerge from the added magnitude constraint.
            
            \par
            \indent Although the new constraint sufficiently corrects for surface brightness bias within our sample a new issue arises that is two-fold. As shown in the right panel of Figure \ref{SB}, this limit excludes a large portion of non-clumpy SFGs from our sample which affects completeness from the inclusion of only bright SFGs. This also creates a sample of clumpy and non-clumpy galaxies that are nearly identical, as indicated by the nearly matching distribution medians shown in left column of Figure \ref{results_25_8p}. The concern with these issues is that it may be physical that fainter galaxies are non-clumpy. Therefore, we find it unsuitable to apply this limit to our clump sample as the comparison of clumpy and non-clumpy galaxy properties may be misleading without considering intrinsically fainter systems. However, we note that surface brightness can be an important factor in demographic results.

             % Clump Definition plot placed here for cosmetics in the pdf %
            \begin{figure}[htb!]
                \centering
                \vspace{0.5cm}
                \includegraphics[scale=0.56]{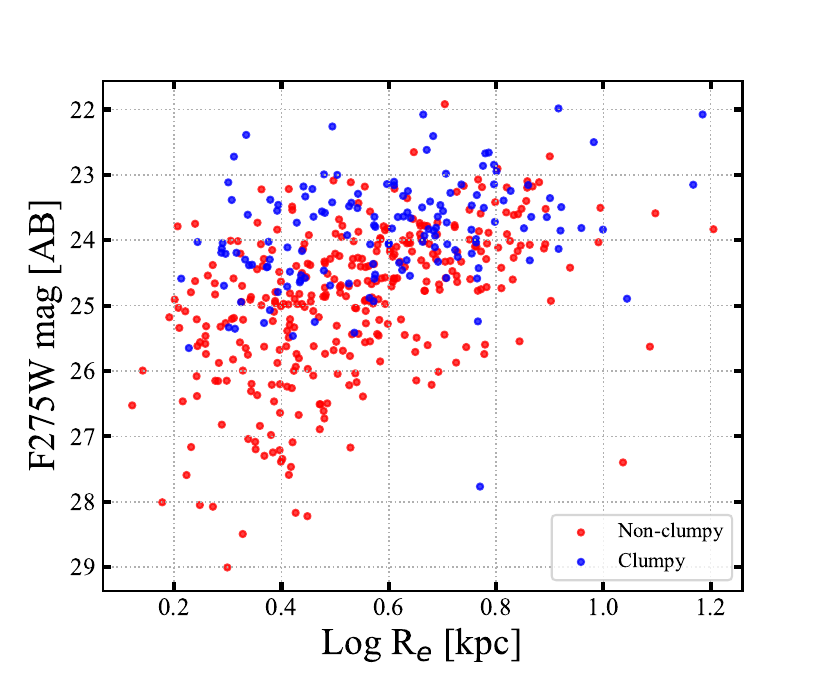}
                \vspace{0.5cm}
                \includegraphics[scale=0.56]{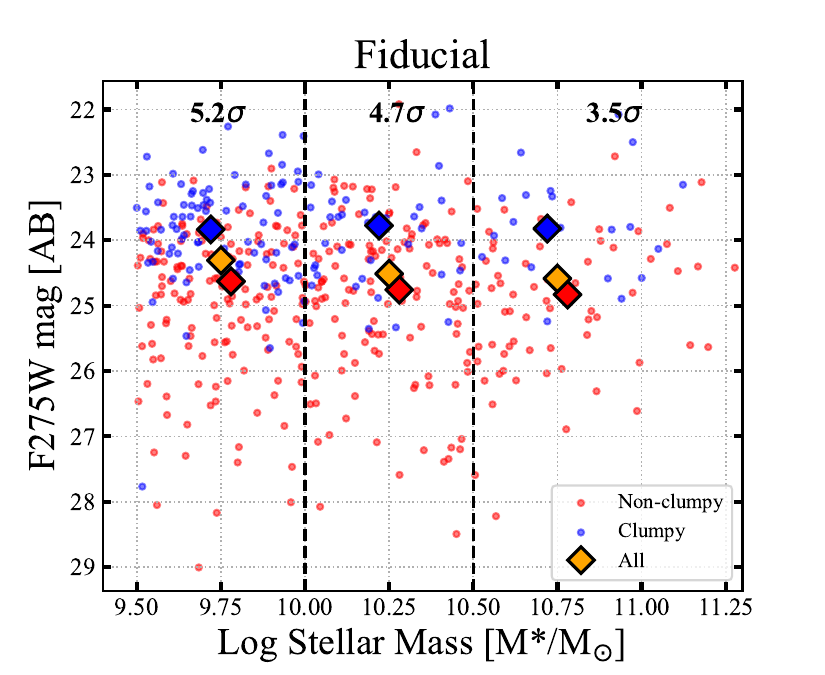}
                \vspace{0.5cm}
                \includegraphics[scale=0.56]{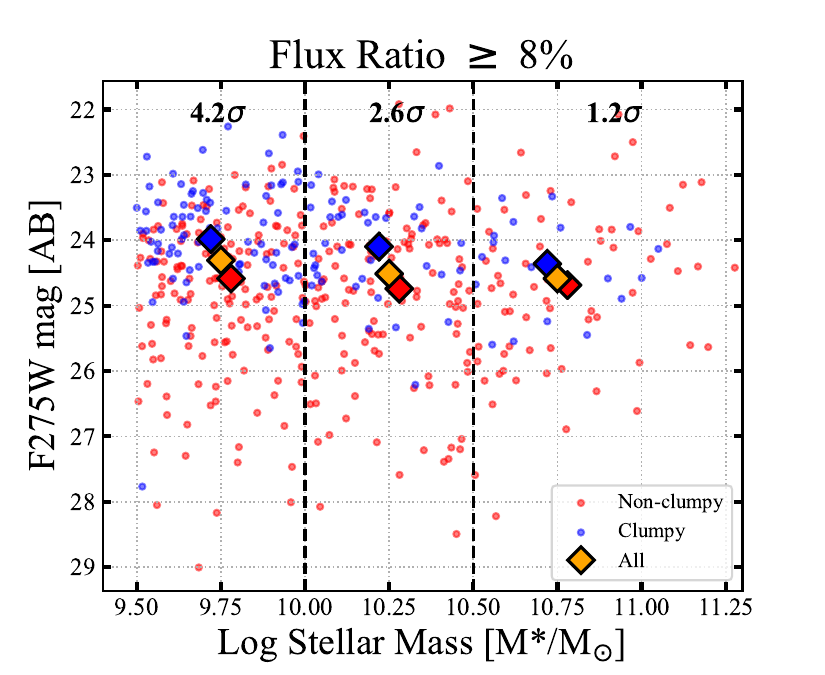}
                \caption{The top panel shows the apparent F275W $-$ Size distribution for our sample, while the middle and lower plots show apparent F275W $-$ Mass relations. The middle plot represents our fiducial SFG sample with individual clump luminosity, M$_{UV} \leq -$16 AB while the bottom plot represents a sample of flux ratio $\geq$ 8\%. The structure of each panel is the same as in Figure \ref{results}.}
                \label{f_8_mag}
            \end{figure}

             % % % % Star forming plots placed here for cosmetics % % % %
            \begin{figure*}[htb!]
                \centering
                \includegraphics[scale=0.43]{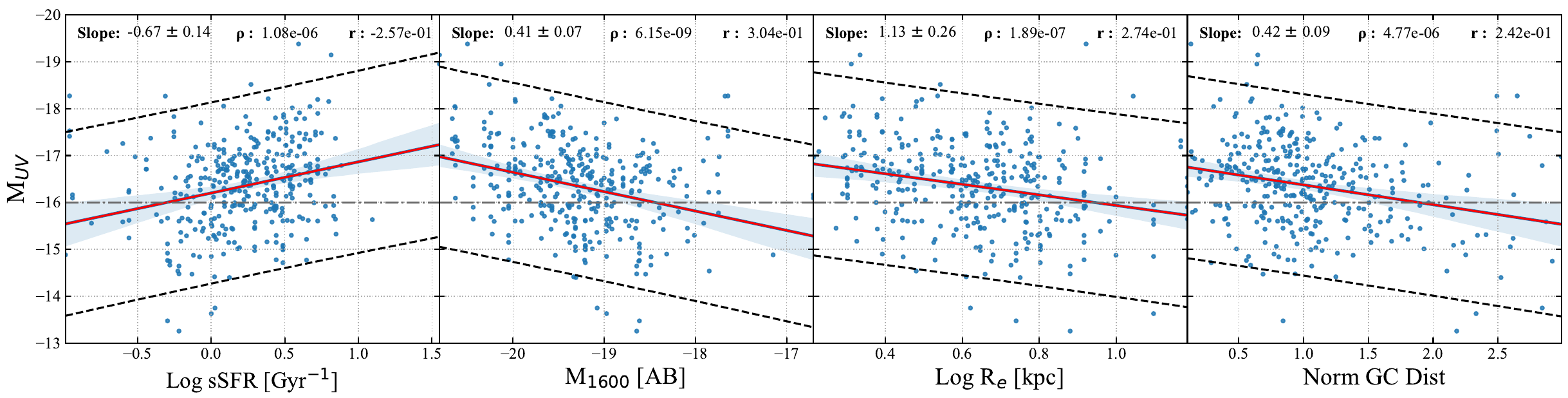}
                \caption{Clump luminosity (expressed in UV absolute magnitude) versus host galactic properties for our UVCANDELS clump distribution. The galactic properties (x-axis) shown in panels 1$-$3 are the host galaxy's specific star-formation rate, absolute magnitude in the rest-frame 1600 \AA, and size. Panel 4 shows the clump's galactocentric distance normalized by the host galaxy's size. The clumps (blue points) are fitted linearly (red line) to show the robustness of each of the gradients. The blue shaded region and the dashed black lines indicate the 95\% confidence and prediction intervals respectively. At the top of each plot we show the Spearman's rank coefficient ({$\rho$}- and r- values) statistics of each relation.}
                \label{lum_plots}
            \end{figure*}  
    
        \subsubsection{Clump Luminosity Threshold} \label{cft}
            
            \par
            \indent For our clump definition we define clumps based on a luminosity minimum (M$_{UV} \leq -$16 AB) that separates them from typical (local) HII regions as described in Section \ref{photometry}. However, as mentioned, the resulting clump sample still may suffer from incompleteness. Here we compare three different luminosity thresholds (no constraint, M$_{UV}\leq -$16 AB, and M$_{UV}\leq -$16.7 AB) to show the impact on our galactic property relations.

            \par
            \indent Figure \ref{cft_all}, located in the appendix, shows the effects of each luminosity threshold on each of our 4 galactic property results. From left-to-right we varied the threshold with no added constraint (`all' clumps), our fiducial result, and $-$16.7 Mag which is less affected by incompleteness in our clump sample.
            %which is corrected for incompleteness in our clump sample.

            \par
            \indent For galaxy size we find size dependence only within our `all' clump sample for intermediate and high mass SFGs and within our fiducial sample for high mass SFGs. Our M$_{UV}\leq -$16.7 AB constraint shows no size dependence in all mass bins. For galaxy SFR, we find distinction between clumpy and non-clumpy SFGs for all constraints in low mass SFGs and within our `all' clump and fiducial samples for intermediate mass SFGs. We find no distinction in each constraint for high mass SFGs. For galaxy color we find distinction in all mass bins for all constraints, with the exception of our fiducial and M$_{UV} \leq -$16.7 AB samples for high mass SFGs. Lastly, we find no change in galaxy morphology with each constraint as the overall structure related to the S\'ersic index remains `disk-like' for all mass bins of each population. We note the large increase in S\'ersic index to high mass clumpy SFGs, which most likely arises from low statistics for our fiducial and complete samples. 

            \par
            \indent We conclude that our threshold definition has a small effect on our results due to the disagreement in galaxy SFR between our complete and fiducial sample for intermediate mass SFGs. However, we argue that the discrepancies seen within high mass galaxy size can be explained by small-number statistics and that our overall results for our fiducial sample (compared to the complete sample) are robust. We do emphasize that it is necessary to distinguish clumps from HII regions as applying no constraints has a major impact on galaxy size dependence.
            
        \subsubsection{Clump Definition}
            
            \par
            \indent Clumps, and therefore clumpy galaxies, have been previously defined differently using minimum thresholds to the luminosity ratio of the brightest detected clumps within each galaxy \citep{2014ApJ...786...15M} and their total UV contributions \citep{2015ApJ...800...39G,2016ApJ...821...72S,2020MNRAS.499..814H,2022ApJ...931...16A}. In this section we discuss how our defined clump luminosity threshold sample compared to a common definition first presented in \citep{2015ApJ...800...39G} who required that individual clumps contribute $\geq$ 8\% of the total UV flux of their host galaxies.
            
            \par
            \indent The right column of Figure \ref{results_25_8p}, located in the appendix, shows demographic results for our SFG sample with clumps defined as having $\geq$ 8\% of the total UV flux of their host-galaxies. We find that the results fully match our complete sample (M$_{UV} \leq -$16.7 AB) results from Section \ref{cft}. Again we find no size or morphology dependence for all mass bins and only a $\geq$3$\sigma$ for low mass galaxy SFR as well as low and intermediate mass galaxy color. We note that the applied constraint changed from an intrinsic property (clump luminosity) for our fiducial sample, to a minimum flux ratio in which the brightness of the galaxy is a significant factor, as a brighter host galaxy causes a higher clump brightness threshold. It is also known that the size and brightness of a galaxy is directly connected with larger galaxies being brighter on average shown in the top panel for our own sample in Figure \ref{f_8_mag}. Therefore, we investigate directly the changes to the UV apparent magnitude between the two constraints shown in the middle and bottom panel of Figure \ref{f_8_mag}.
            
            \par
            \indent As indicated by the bottom panel (Figure \ref{f_8_mag}), applying the 8\% flux ratio minimum switches some of the brightest clumpy galaxies within each mass bin to a non-clumpy classification due to a higher clump brightness threshold redefining their clumps. Compared to our fiducial results, this reclassification directly causes the decrease/increase in medians seen for clumpy/non-clumpy galaxies within our 8\% minimum size results in the top panel of Figure \ref{results_25_8p}. Furthermore, higher star-formation results in higher UV luminosity from a SFG, and a change to the brightest UV galaxies would also explain the decrease/increase in medians seen in the SFR high mass bin (second panel of Figure \ref{results_25_8p}). We conclude that both constraints (fiducial and flux ratio $\geq$ 8\%) provide robust results but lead to different conclusions regarding properties that depend on the host galaxy’s brightness. We leave it to future studies to decide whether or not they wish to define their clumps based on intrinsic clump properties (fiducial) or overall galaxy properties (flux ratio $\geq$ 8\%). 
            
            % Within each mass bin a portion of the brightest clumpy galaxies switch to non-clumpy after applying the 8\% flux ratio minimum. This effect directly causes the decrease/increase in medians seen for clumpy/non-clumpy galaxies within our 8\% minimum size results. Furthermore, higher star-formation results in higher UV luminosity from a SFG and a change to the brightest UV galaxies would also explain the decrease/increase in medians seen in the SFR high mass bin. Therefore, both constraints (fiducial and flux ratio $\geq$ 8\%) provide robust results, but lead to different conclusions regarding properties that depend on the host galaxy's brightness. We leave it to future studies to decide whether or not they wish to define their clumps based on intrinsic clump properties (fiducial) or overall galaxy properties (flux ratio $\geq$ 8\%).    
    \subsection{Star-forming Region Luminosity - Galaxy Properties Relation} \label{cfvgp}
        \indent As described in detail in Section \ref{photometry} we measured the rest-frame UV luminosity for each clump detected in our UVCANDELS sample. In Figure \ref{lum_plots} we directly compare our clump luminosities to several distinct galactic properties (host sSFR, luminosity, size, and normalized clump galactocentric distance). Here we include all detected star-forming regions beyond our luminosity limit of M$_{UV}$ $\leq -$16 AB to provide a comprehensive picture of star-forming regions and host galaxy relations independent of our clump definition. Subsequently, this inclusion also increases the dynamical range of the investigated star-forming region luminosities. Each relation was fit linearly to determine a gradient for our distribution of clumps. We also show the Spearman's rank coefficient ($\rho$- and r- values), which denote the statistical dependence between the rankings of two variables, at the top of each panel to demonstrate that the relations seen are significant beyond a linear fit. For these properties we find robust relations ($\geq$ 3$\sigma$) indicating that on average more luminous clumps reside in small, but luminous SFGs that have higher sSFR. Furthermore, the right-most panel of Figure \ref{lum_plots} indicates that more luminous clumps are found closer to the galactic center of each SFG, regardless of the host's overall size. 
          
\section{Discussion}
\label{Discussion}

\subsection{What Types of Galaxies Tend to Form Clumps?}
\label{Discussion:galaxies}

%\indent Generally, two of our main results are (1) clumpy galaxies have higher SFR (and bluer restframe U-V colors) than non-clumpy galaxies and (2) clumpy galaxies are only marginally larger than non-clumpy galaxies. These two results are supporting the scenario of clumps being formed through VDI. 

% My intro paragraph revising the above one
\indent Based on our results, our main conclusions are that (1) clumpy galaxies exhibit higher star formation rates (SFR) and appear bluer in restframe U-V colors compared to non-clumpy galaxies and (2) clumpy galaxies are only marginally larger than non-clumpy galaxies. These two results provide support for the hypothesis that clumps form through violent disk instability (VDI).

\par 
\indent First, the systematic difference between median SFR of clumpy and non-clumpy galaxies is inconsistent with the prediction from major mergers. The 2nd panel of Figure \ref{results} shows that in low/intermediate/high regimes, clumpy galaxies have higher SFR by a factor of 2, 2 and 1.7 respectively to that of non-clumpy galaxies. This difference is larger than the difference caused by mergers. According to \citet{2019A&A...631A..51P}, merging galaxies in CANDELS do not significantly affect the SFR of galaxies compared to non-merging galaxies. Their findings indicate a systematic increase of SFR caused by merger is only a factor of 1.2. Moreover, minor mergers are not able to result in such a large observed SFR difference as well.
% % % Reworded this commented section below % % %
% We assume each minor merger component could contribute an additional SFR up to a few tens percents to its host galaxy. Given the average clump UV luminosity compared to the host galaxies' UV luminosity (see also \citet{2015ApJ...800...39G} and \citet{2016ApJ...821...72S}), to increase the host SFR by a factor of $\sim$2 requires having AAA/BBB/CCC minor merger components at the same time. --- {\bf How am I supposed to calculate this?} --- These numbers are much larger than the average clump numbers per galaxy in our sample (Figure \ref{Cnum}). 
We assume each minor merger component could contribute an additional SFR up to a few tens percents to its host galaxy considering that the average clump UV luminosity compared to the host galaxies' UV luminosity is $\sim$6.6\% for our sample (see also \citet{2015ApJ...800...39G} and \citet{2016ApJ...821...72S}). Assuming a generous boost of 10\% to host SFR with each minor merger, to increase the host SFR by a factor of $\sim$2 requires having 10 minor merger components at the same time. This number is much larger than the average clump numbers per galaxy in our sample (Figure \ref{Cnum}).
Therefore, we argue that the clumpy appearance and high SFR are associated with some internal reasons.

\par
\indent This scenario is consistent with in-situ formation through VDI. To form clumps in a low-redshift unstable disk, the Toomre $Q$ parameter should be less than unity\footnote{See, however \citet{2016MNRAS.456.2052I} for the situations at high redshift.} \citep{1964ApJ...139.1217T,2007ApJ...658..763E,2009ApJ...701..306E,2007ApJ...670..237B,2009ApJ...707L...1B,2008ApJ...687...59G,2011ApJ...733..101G,2009ApJ...703..785D,2022MNRAS.511..316D,2009MNRAS.397L..64A,2010MNRAS.404.2151C}. Here we use the formulas of \citet{2011ApJ...733..101G} to investigate the dependence of $Q$ on SFR and $R_e$. We start from Equation (3) of \citet{2011ApJ...733..101G},
\begin{equation}
    Q_{gas} = \left( \frac{\sigma_0}{V_c} \right) \left( \frac{a(V_c^2R_d)/G}{\pi R_d^2 \Sigma_{gas}} \right),
    \label{eq1}
\end{equation}
where $\sigma_0$, $V_c$, $R_d$, $\Sigma_{gas}$ are the velocity dispersion, circular velocity, disk radius, and cold gas surface density of a disk, respectively. For simplicity, we only discuss the gas component here and ignore all the constants (geometric coefficient $a$, $\pi$, and $G$). To connect $\Sigma_{gas}$ to SFR, we use the inverted Schmidt-Kennicutt Law (see Equation (2) of \citet{2011ApJ...733..101G}): $\Sigma_{gas} \propto (\Sigma_{SFR})^\alpha \propto (SFR/R_d^2)^\alpha$, where $\alpha = 0.73$. We also have $V_c \propto \sqrt{M_{tot}/R_d}$, where $M_{tot}$ includes the mass of stars, gas, and dark matter. According to \citet{2016ApJ...831..149W} the baryonic budget, $log(M_{*}/M_{tot}) \sim 0.58$ at $z\sim 1.0$, with a small scatter of about 0.2 dex. For simplicity, we assume the scatter is caused mainly by $M_{*}$ dependence (e.g., see Figure 5 of \citet{2016ApJ...831..149W}) rather than by $R_e$ dependence. Therefore, in a given $M_{*}$ bin, we assume $M_{tot}$ has no correlation with $R_e$. We also assume $R_d \propto R_e$, the effective radius. This assumption is valid because on average, the galaxies in our sample have Sersic index $n \sim 1$ (the bottom panel of Figure \ref{results}). For this profile, the radius containing 95\% of galaxy light is about 3$-$4 $R_e$ (See \citet{2005PASA...22..118G} for a concise reference of Sersic R$^{1/n}$ profiles).

\par 
\indent With all above assumptions, we have,
\begin{equation}
    Q \sim Q_{gas} \propto \sigma_0 M_{tot}^{1/2} R_e^{2\alpha-3/2} (1/SFR)^{\alpha}.
    \label{eq2}
\end{equation}

Given that $\alpha = 0.73$, at a fixed $\sigma_0$ and $M_{tot}$, we have $Q \propto R_e^{-0.04} SFR^{-0.73}$. This result suggests that the higher the SFR a galaxy has, the lower its $Q$, and therefore the easier for the galaxy to form clumps. Meanwhile, $Q$ only shows little, if any, dependence on $R_e$. Even if we considered a mild correlation between $\sigma_0$ and $R_e$ (for example, $\sigma_0 \propto R_e^{-0.14}$, see \citet{2013ApJ...767..104N}), the dependence of $Q$ on $R_e$ would still be marginal $Q \propto R_e^{-0.18}$. This deduction is in very good agreement with the top and second panel of Figure \ref{results}. Therefore, we argue that our results provide strong supporting evidence for VDI.

\subsection{How Luminous are the Clumps Formed in Different Galaxies?} \label{Discussion:clumps}

\indent Another major result of this paper is the correlations between the UV luminosity of star-forming regions and host galaxy properties (Figure \ref{lum_plots}). Among these relations, the most interesting one is the negative correlation between clump luminosity and their host galaxy size. Since galaxy size increases with their $M_*$ \citep{2014ApJ...788...28V}, to remove the mass dependence, we also study this clump luminosity -- galaxy size correlation in our three mass bins separately. The negative correlation persists in the low- and intermediate-mass bins. The high-mass bin, however, shows no correlation, likely due to the small sample size and poor number statistics. Here we discuss the implications of this result.

\par 
\indent According to \citet{2008ApJ...685L..31E}, the mass of clumps formed through VDI can be calculated as 
\begin{equation}
    M_{cl} \propto M_{gas} \eta^2 = M_{gas} \left( \frac{M_{gas}}{M_{tot}} \right)^2.
    \label{eq3}
\end{equation}
Similar to the discussion above, we have $M_{gas} \propto \Sigma_{gas} R_e^2 \propto \Sigma_{SFR}^\alpha R_e^2 \propto SFR^\alpha R_e^{2-2\alpha}$ and $M_{tot}$ is not correlated with $R_e$ at a given $M_*$. We then have
\begin{equation}
    M_{cl} \propto SFR^{3\alpha} R_e^{3(2-2\alpha)}.
\end{equation}
This relation predicts that the maximum clump mass increases with two parameters: (1) SFR and (2) $R_e$, when $\alpha = 0.73$. The first correlation with SFR is supported by panel 1 and 2 of Figure \ref{lum_plots}, if we assume clumps are simple stellar population systems so that their $M_{UV}$ is a good representative of their $M_*$. The second correlation with $R_e$, however, is opposite to our observed result.

\begin{figure}[htb!]
        \centering
        \includegraphics[scale=0.65]{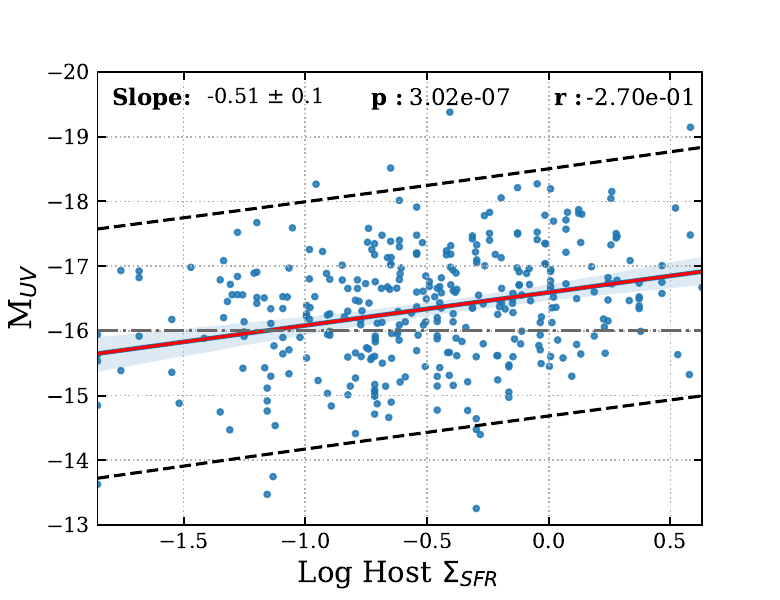}
        \caption{Clump luminosity versus host galactic SFR density. The format is the same as in Figure \ref{lum_plots}.}
        \label{clum_SFRden}
        \end{figure}

This inconsistency could have a few explanations. First, at face value, it suggests that there is another parameter that determines clump mass rather than $R_e$. We speculate this parameter is $\Sigma_{SFR}$ (or $\Sigma_{gas}$). Indeed, the clump luminosity shows a correlation with $\Sigma_{SFR}$ as depicted in Figure \ref{clum_SFRden}. $\Sigma_{gas}$ also shows a similar trend if we convert $\Sigma_{SFR}$ to $\Sigma_{gas}$ by using the Kennicutt-Schmidt Law. Since $\Sigma_{SFR}$ is inversely proportional to $R_e^2$, this speculation is consistent with the first three panels of Figure \ref{lum_plots}. Second, in our discussion above, we use clump UV luminosity $M_{UV}$ to represent clump mass $M_{cl}$. This simplification would not be valid if clumps have different stellar populations and dust extinction. Therefore, deriving clump mass $M_{cl}$ would provide a more straightforward comparison to models. This work will be presented in a future paper. Third, related to the issue of $M_{cl}$, Figure \ref{lum_plots} in fact shows a mixed result of clump formation and evolution. Its fourth panel shows a gradient of clump luminosity with respect to their normalized galactocentric distance. This result is consistent with the inward migration scenario of clump evolution \citep{2007ApJ...670..237B,2008ApJ...688...67E,2009ApJ...703..785D,2022MNRAS.511..316D,2010MNRAS.404.2151C,2011ApJ...739...45F,2014MNRAS.443.3675M,2014ApJ...780...57B,2018ApJ...853..108G,2016ApJ...821...72S,2019MNRAS.489.2792Z,2021MNRAS.506.3916L}. Therefore, it is possible that the most luminous clumps (i.e., those found in small galaxies) are also those which have already migrated close to the center of their galaxies. At a given $M_*$, small galaxies have higher mass density and therefore shorter dynamical timescale, which would result in a faster clump migration in small galaxies than in large galaxies. During this fast migration, merger or continuing growth of clumps \citep{2007ApJ...670..237B,2014ApJ...780...57B,2009ApJ...701..306E,2009ApJ...703..785D,2022MNRAS.511..316D} would increase both clump mass and UV luminosity. This scenario is supported by the locations of the most luminous clumps in the second and fourth panels of Figure \ref{lum_plots}. To further shed light on this scenario as well as the maximum $M_{cl}$, the age of clumps needs to be determined. We will present relevant work in a future paper.

\section{Summary and Conclusion}
\label{Conclusion}
    
    \par
    \indent In this work we used the newly obtained F275W 60-mas mosaics from UVCANDELS for 4 of the premier CANDELS fields (GOODS-N, GOODS-S, EGS and COSMOS) to detect our sample of UV-selected clumps at 0.5 $\leq$ z $\leq$ 1, thus sampling their rest-frame 1600 \AA. By subdividing our host SFGs into `clumpy' and `non-clumpy' galaxies based on whether or not they contain at least 1 off-center clump within their structure, we determined different distinctions for host galaxy properties as well as clumpy luminosity relations. Our final fiducial sample applies a luminosity threshold to individual clumps, requiring them to be 3 times brighter in their UV emission than the most massive/brightest HII regions found within the local universe. In doing so we shed light on the physical properties of galaxies that host clumps and what distinguishes them from the general population of star-forming galaxies. Furthermore, we have shown several effects on our results of sample selection and constraints that are commonly used throughout the literature to provide a guide towards potential bias in our clump samples for future work. Our major results are listed as follows:
    
    \begin{itemize}
        \item We present the clumpy fraction (the fraction of clumpy galaxies with respect to the total) of our sample of SFGs in Figure \ref{CF}. We find that for low mass galaxies $f_{clumpy} \sim$ 35\% which decreases to $\sim$ 22\% for the intermediate and high mass bins. This result shows that clumpy SFGs make up a significant fraction of all SFGs at all galaxy masses at 0.5 $\leq$ z $\leq$ 1 which is consistent with other UV-selected samples found within the literature.
        
        \item Using a distinction limit of ($\geq$3$\sigma$) we find that clumpy SFGs on average have higher SFR and are bluer in rest-frame U-V color for low and intermediate masses; and are larger in size for massive SFGs when compared to non-clumpy SFGs of the same mass. Clumpy SFGs are, however, similar in overall morphology to non-clumpy SFGs as both clumpy and non-clumpy SFGs contain disk-like structures. Each of our results supports the scenario of in-situ clump formation through VDI.
        
        \item For the distributions of clump luminosity vs galactic properties we find that all detected clumps show robust linear relations for host sSFR, UV-luminosity, size and clump galactocentric distance. We conclude that the relations shown indicate inward migration of clumps through their host disk profiles, however, future work deriving the clump ages is needed to further confirm this theory.
        
        \item We test the effect of different galaxy selections and clump definitions on the demographics of clumpy galaxies. The overall result is summarized in Figure \ref{sigma_SS} in the Appendix. In each panel with 3$\times$3 cells, if any of the four surrounding cells shows a different color from the central cell, the demographic result is changed. Overall, the summary is (1) The demographics of low-mass galaxies are not changed by various galaxy or clump selections; (2) The lack of difference in Sersic index between clumpy and non-clumpy galaxies are robust; (3) Using UVJ or sSFR to select SFGs generally does not affect the demographics; (4) Using a more restrained galaxy or clump selection method to ensure high completeness would reduce the sample size as well as reduce the difference between clump and non-clumpy galaxies (namely, the more conditions one applies, the more similar the two selected populations are).

    \end{itemize}
    
    \par
    \indent In future work, we will explore two avenues that will build upon these results. The first project will determine clump properties through SED fitting of the rest-frame FUV(F275W) to mid-IR(F160W) passbands and dust extinction estimates. We aim to determine the most accurate estimates of masses, ages, star-formation rates (both UV-corrected and H$\alpha$ based) and burstiness for the largest sample of clumps to date. In doing so, we will utilize the ages and established property gradients to shed light on the longevity of clumps as well as possible evolutionary tracks of clumps as galaxies evolve over time. Secondly, we will compare our results with additional observations out to z $\sim$ 3 with an ensemble of simulations with varying feedback levels in order to determine maximum constraints on theoretical modeling of galaxies. In brief, the clumpy nature of these substructures can be smoothed out of simulated galaxies if excessive levels of feedback (e.g. UV/IR radiation pressure, supernovae, stellar winds, etc.) are implemented within the models. As we have observed, primordial galaxies host clumps in their morphologies making it vital that our constraints to feedback reproduce these same clumpy structures within our theoretical models.
    
    \begin{center}
        \large{Acknowledgements}
    \end{center}
    
    \par
    \indent This work is based on observations with the NASA/ESA Hubble Space Telescope obtained at the Space Telescope Science Institute, which is operated by the Association of Universities for Research in Astronomy, Incorporated, under NASA contract NAS5-26555. Support for Program numbers HST-GO-15647 and HST-AR-15798 was provided through a grant from the STScI under NASA contract NAS5-26555.
    \par 
    \indent DC is a Ramon-Cajal Researcher and is supported by the Ministerio de Ciencia, Innovaci\'{o}n y Universidades (MICIU/FEDER) under research grant PGC2018-094975-C21.
    \par
    \indent YSD acknowledges the support from National Key R\&D Program of China for grant No.\ 2022YFA1605300, and the NSFC grants No. \ 12273051 and 11933003.
    \par
    \indent The specific observations analyzed can be accessed via \dataset[10.17909/96zb-g146]{http://dx.doi.org/10.17909/96zb-g146}.

\appendix
        
    \section{Extensive Review of Sample Selection Effects}

        \begin{figure*}[htb!]
            \centering
            \includegraphics[scale = 0.72]{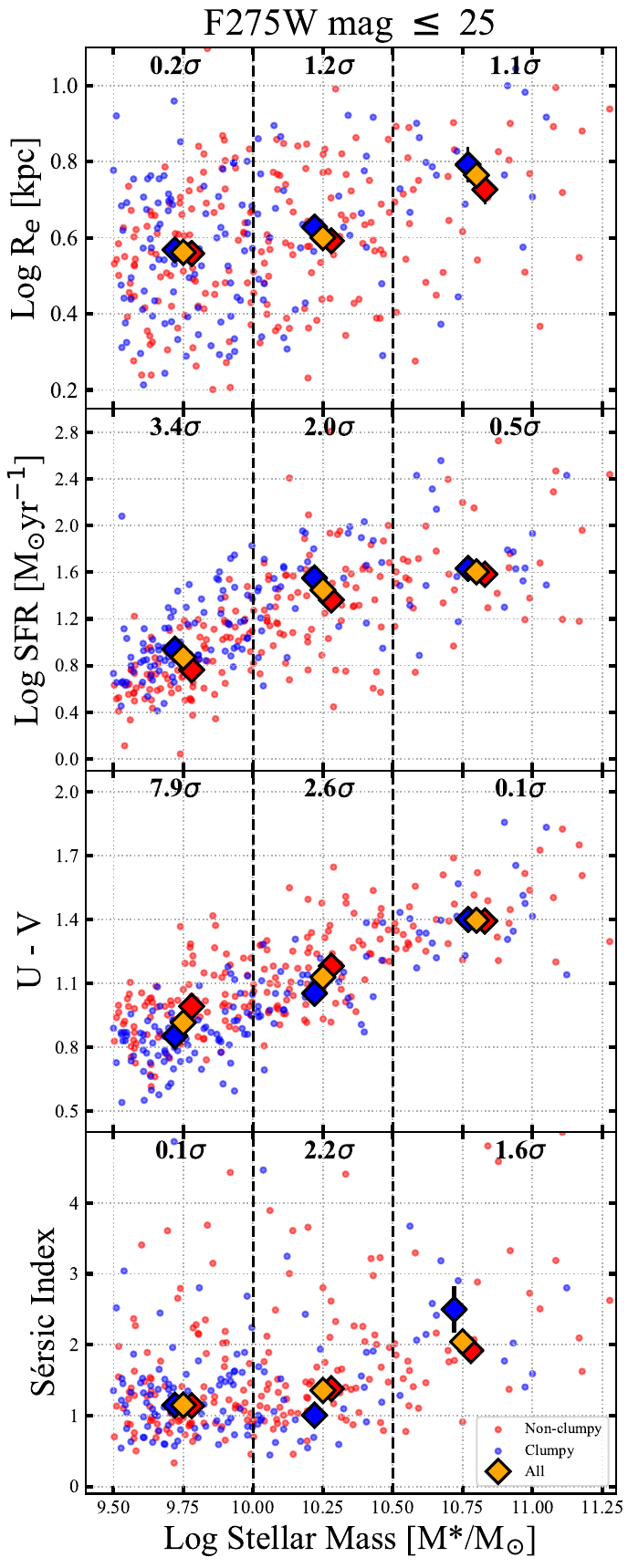}
            \includegraphics[scale=0.72]{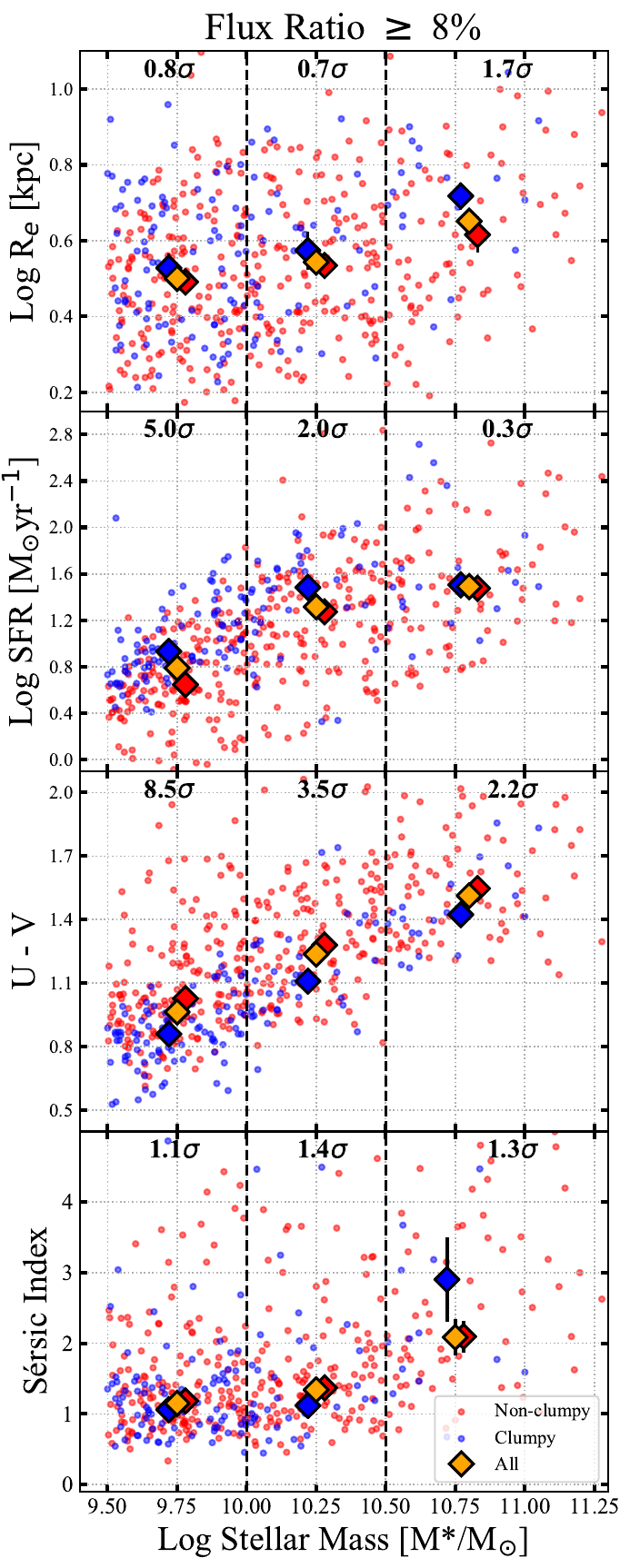}
            \caption{Relations of host-galaxy properties versus stellar mass for clumpy and non-clumpy galaxies with magnitude F275W $\leq$ 25 AB (left) and whose clumps contribute at least 8\% of the total UV luminosity. From top to bottom, we show host-galaxy size, SFR, color and morphology. Symbols, colors and lines are the same as in Figure \ref{results}.}
            \label{results_25_8p}
        \end{figure*}
            
        \par
        \indent Here we display the plots that indicate the effects of sample selection on our results for our four galactic properties $-$ galaxy size, SFR, color and S\'ersic index. Figure \ref{results_25_8p} shows (left) the demographic results for our galaxy brightness constraint of m$_{UV} <$ 25 mag to correct for surface brightness bias. As described in Section \ref{sb}, we find this has a major effect on our results as the medians of each mass bin for each galactic property become nearly identical. The right side of Figure \ref{results_25_8p} shows the demographic results for clumps that provide at least 8\% of their host galaxy's total UV luminosity.
        
        \par
\begin{figure*}[htb!]
            \centering
            \includegraphics[scale = 0.73]{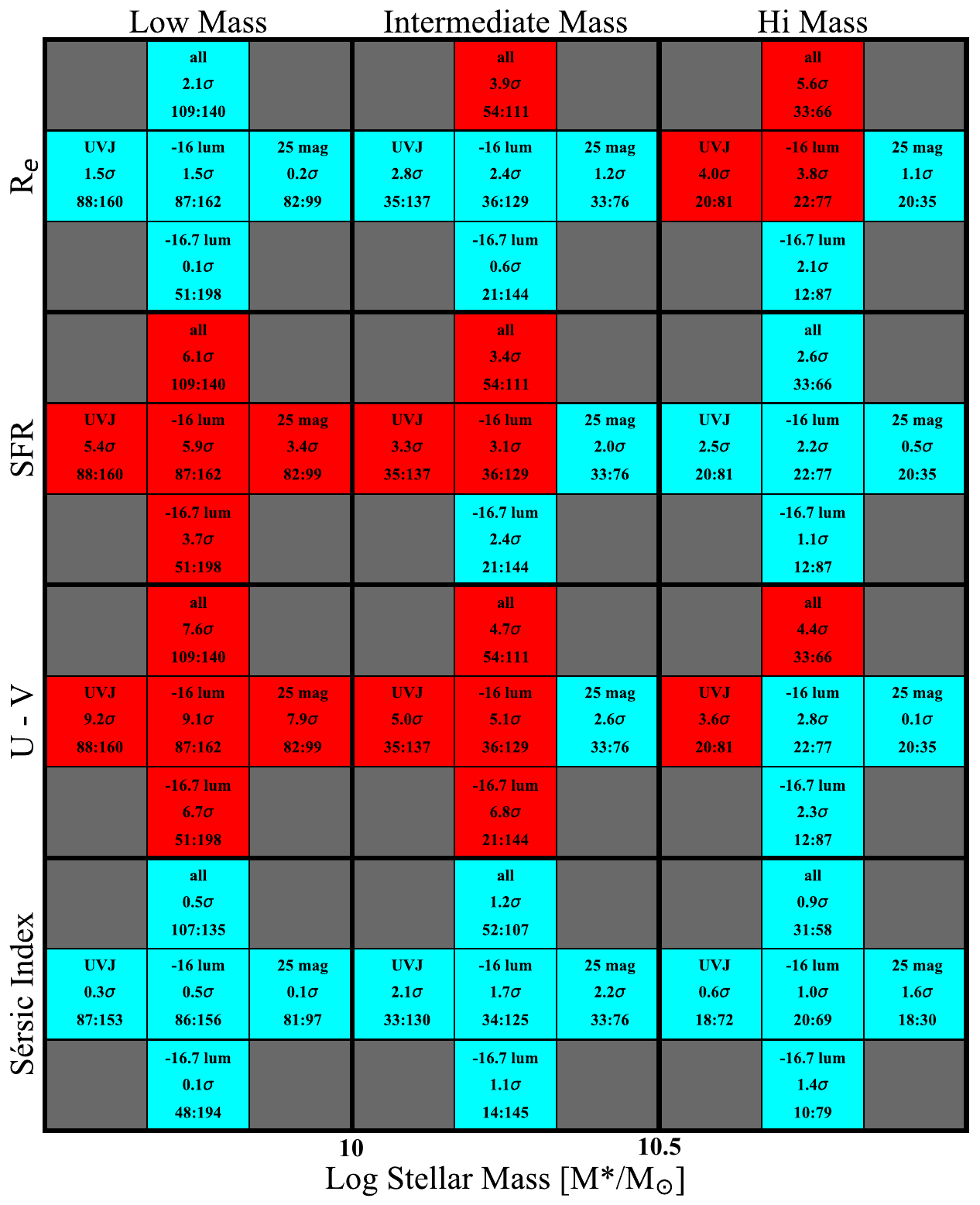}
            \caption{Complete representation of all sample selection effects investigated in this work. Namely, we test the effects of including all clumps; selecting clumps above a threshold luminosity of M$_{UV} \leq -$16.0 or $-$16.7 AB (see Figure \ref{flim}); defining SFGs using the UVJ color$-$color method; and our 25 UV magnitude correction for surface brightness. The overall structure is divided into 12, 3$\times$3 panels representing the effects of different SFG and clump selection criteria on the four host-galaxy properties studied in this work $-$ galaxy size, SFR, color, and S\'ersic index $-$ in our three mass bins. Within each 3$\times$3 panel there are 5 pixels which refer to the 5 different selection criteria for SFGs/clumps. These are named in the relevant pixels, along with the statistical deviation between clumpy and non-clumpy SFGs in the relevant property, and the ratio of clumpy to non-clumpy galaxies. Red squares indicate statistically significant deviations between the two populations, while cyan squares indicate they are statistically similar.}
            \label{sigma_SS}
        \end{figure*}

        \indent In Figure \ref{sigma_SS} we show a comprehensive representation of all of our sample selection comparisons between clumpy and non-clumpy SFGs performed throughout this work. We again show the comparison for galaxy host size, SFR, color and morphology and represent the two populations as distinct/similar by coloring them as red/cyan squares respectively. Within this figure we do not show the results for our 8\% flux ratio sample as it produced the same conclusions as our $-$16.7 Mag threshold sample. Therefore, we only represent the luminosity threshold results within Figure \ref{sigma_SS}.
        
        \section{Clump Luminosity Threshold Plots}

            \begin{figure*}[htb!]
                \centering
                \includegraphics[scale=1.02]{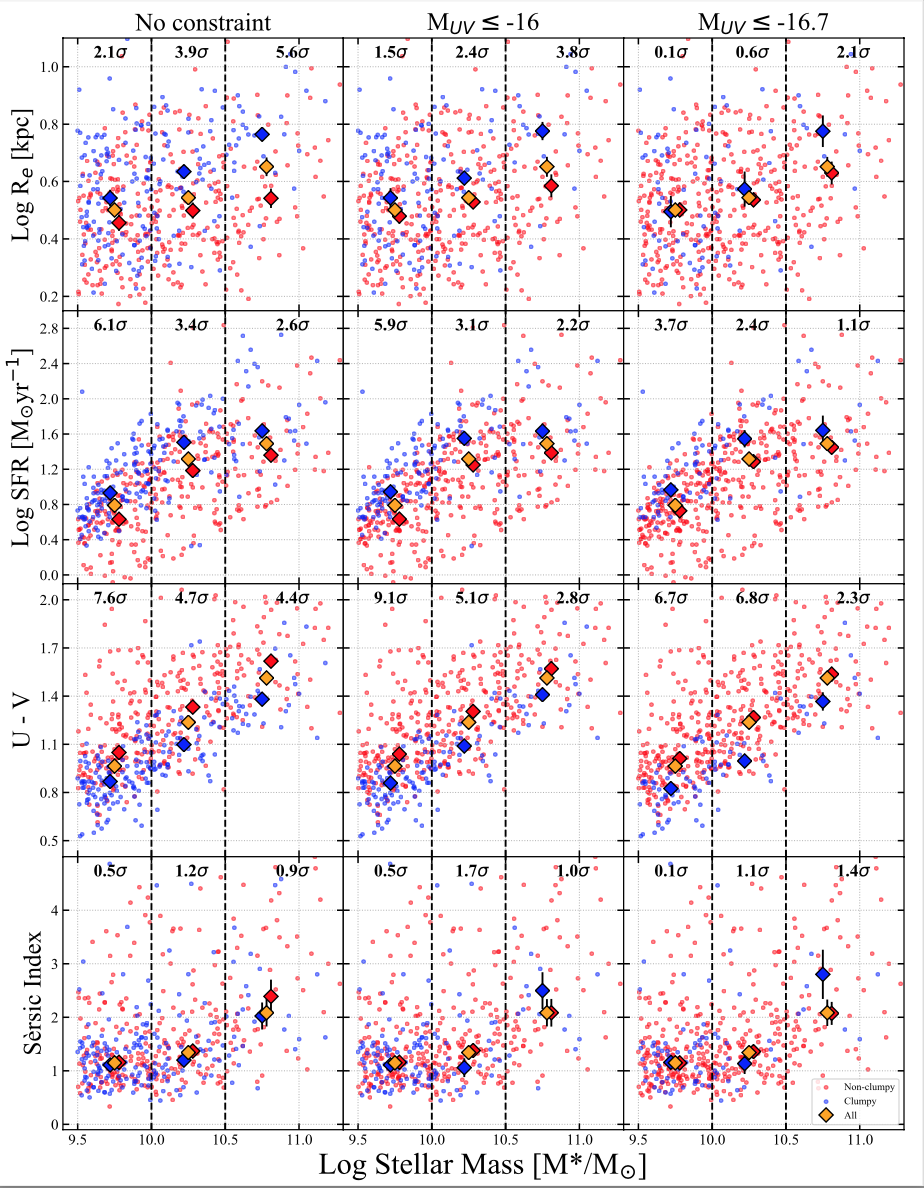}
                \caption{Galactic Property $-$ Stellar Mass results for host size, SFR, color and morphology. The structure is the same as Figure \ref{results}. The different columns represent the different minimum clump luminosity thresholds used for our analysis (indicated at the top of each respective column).}
                \label{cft_all}
            \end{figure*}
        
            \indent In Figure \ref{cft_all} the results for the different clump luminosity thresholds for our four galactic properties $-$ galaxy size, SFR, color and S\'ersic index $-$ with (no constraint, M$_{UV} \leq -$16 AB, and M$_{UV} \leq -$16.7 AB). We find that including `all' detected clumps within our sample reveals differences between our two populations (in low and intermediate mass galaxy size) not present in our fiducial sample. However, the inclusion of all detections does not distinguish the sample of clumps from typical HII regions. Differences in conclusions for our strict M$_{UV} \leq -$16.7 AB constraint to our fiducial sample for galaxy size may result from low number statistics of clumpy galaxies as described in the main body of the text.

    \bibliography{References}
    \bibliographystyle{aasjournal}

\end{document}